\documentclass[superscriptaddress,aps,twocolumn,prd,showpacs,altaffilletter,nofootinbib]{revtex4-1}
\usepackage{amsmath}
\usepackage[utf8]{inputenc}
\usepackage{amsfonts}
\usepackage{amssymb}
\usepackage{graphicx}
\makeatletter
\newcommand*{\rom}[1]{\expandafter\@slowromancap\romannumeral #1@}
\makeatother

\usepackage[usenames,dvipsnames]{xcolor}
\usepackage{hyperref}
\hypersetup{colorlinks=true,
citecolor=MidnightBlue, linkcolor=MidnightBlue, urlcolor=MidnightBlue}
\usepackage{tikz}

\usepackage{comment}

\newcommand{\be}{\begin{equation}}
\newcommand{\ee}{\end{equation}}
\newcommand{\bea}{\begin{eqnarray}}
\newcommand{\eea}{\end{eqnarray}}

\newcommand{\bseq}{\begin{subequations}}
\newcommand{\eseq}{\end{subequations}}

\allowdisplaybreaks[1]
\usepackage{amsmath}

\usepackage{needspace}

\usepackage[normalem]{ulem}

\newcommand{\orcidicon}{%
	\begin{tikzpicture}
	\draw[lime, fill=lime] (0,0)
		circle [radius=0.16]
		node[white] {{\fontfamily{qag}\selectfont \tiny ID}};
	\draw[white, fill=white] (-0.0625,0.095)
		circle [radius=0.007];
	\end{tikzpicture}	\hspace{-2mm}
}

\newcommand\orcidIsmael{{\href{https://orcid.org/0000-0002-0606-764X}{\orcidicon}}}
\newcommand\orcidMariam{{\href{https://orcid.org/0000-0002-1529-1889}{\orcidicon}}}
\newcommand\orcidCheYu{{\href{https://orcid.org/0000-0001-8312-759X}{\orcidicon}}}
\newcommand\orcidKostas{{\href{https://orcid.org/0000-0002-0672-1496}{\orcidicon}}}

\begin{document}


\title{New traversable wormholes without invoking exotic matter in \texorpdfstring{$f(\mathbb Q)$} < from the freedom of different connections}


\author{Ismael Ayuso\orcidIsmael\!\!}
\email{ismael.ayuso@ehu.eus}
\affiliation{Department of Physics, University of the Basque Country UPV/EHU, P.O. Box 644, 48080 Bilbao, Spain}
\affiliation{EHU Quantum Center, University of the Basque Country UPV/EHU, P.O. Box 644, 48080 Bilbao, Spain}

\author{Mariam Bouhmadi-López\orcidMariam\!\!}
\email{mariam.bouhmadi@ehu.eus}
\affiliation{IKERBASQUE, Basque Foundation for Science, 48011, Bilbao, Spain}
\affiliation{Department of Physics, University of the Basque Country UPV/EHU, P.O. Box 644, 48080 Bilbao, Spain}
\affiliation{EHU Quantum Center, University of the Basque Country UPV/EHU, P.O. Box 644, 48080 Bilbao, Spain}

\author{Che-Yu Chen\orcidCheYu\!\!}
\email{b97202056@gmail.com}
\affiliation{RIKEN iTHEMS, Wako, Saitama 351-0198, Japan}

\author{Konstantinos Dialektopoulos\orcidKostas\!\!}
\email{kdialekt@gmail.com}
\affiliation{Institute of Space Sciences and Astronomy, University of Malta, Malta, MSD 2080}

\date{\today}

\begin{abstract}
Wormhole solutions in General Relativity can only exist in the presence of exotic matter that violates the null energy condition. In this paper, we investigate the possibility of constructing wormhole solutions in $f(Q)$ gravity in which the extra connection degrees of freedom contribute such that no exotic matter is needed. Considering the Ellis-Bronnikov wormhole as an example, we find that the system admits two families of connections consistent with the spacetime symmetries and the requirements of symmetric teleparallel gravity. For both families, we show that constant non-metricity scalar $Q$ is not merely a simplifying assumption but arises naturally from the field equations when $f_{QQ}$ is regular, and that the resulting exact analytic solutions can satisfy the null energy condition throughout the entire spacetime. For the second family, which possesses richer connection degrees of freedom, we further demonstrate that solutions with a spatially varying $Q$ exist when $f_{QQ}$ is allowed to diverge at a finite value of $Q$, and that these too can satisfy the null energy condition everywhere. Our results show that the nontrivial connection degrees of freedom in $f(Q)$ gravity can naturally be responsible for supporting wormhole geometries, without resorting to any exotic matter.   
\end{abstract}

\maketitle

\section{Introduction}\label{sec:intro}

Wormholes are among the most exotic astrophysical objects ever theorized, and simultaneously some of the most fascinating to the general public. The concept of traveling from one point in the Universe to another through a tunnel—effectively bypassing the constraint imposed by the speed of light—has captured widespread imagination for good reason. In fact, the feasibility of real interstellar exploration may ultimately depend on their existence.

Unfortunately, to date, wormholes remain purely theoretical constructs, with no experimental evidence supporting their existence. In General Relativity (GR), the existence of a traversable wormhole requires the geometry of the spatial slices to satisfy the so-called flaring-out condition at the throat, which ensures that the wormhole opens outward away from its minimum-area surface. For a static spherically symmetric wormhole described by the shape function $b(r)$, this condition is expressed as $b'(r_0)<1$ at the throat $r=r_0$, where $b(r_0)=r_0$ (see Sec.~\ref{sec:wh}). When combined with the Einstein field equations, the flaring-out condition necessarily leads to a violation of the Null Energy Condition (NEC) \cite{Visser:1995cc,bookLobo}, which is defined as
\bea
T_{\mu\nu}k^\mu k^\nu \geq 0\,,
\eea
for any null vector $k^\mu$, where $T_{\mu\nu}$ denotes the stress-energy tensor. Matter violating the NEC is generally referred to as exotic matter, and there is currently no experimental evidence for its existence{{\footnote{Although there is currently no experimental evidence for its existence, it is fully compatible with being the driver of the late-time acceleration of the universe.}}}. Therefore, if such matter does not exist in our Universe, the existence of static wormholes would be ruled out within the framework of GR. This constraint is relaxed for time-dependent wormhole solutions, for which both the NEC and the Weak Energy Condition (WEC) can be satisfied in certain regions and over specific time intervals at the throat \cite{Hochberg:1998ii, Hochberg:1998ha, Kar:1995ss, Arellano:2006ex}. 

The issue that traversable wormholes in GR require the violation of the NEC has motivated the study of modified theories of gravity, in which additional geometrical contributions arising from modifications of the gravitational sector can effectively provide the required support for wormhole solutions, potentially avoiding the need for exotic matter. In general, such modified gravity theories possess greater flexibility, often leading to a richer and more complex wormhole structure.

Many of these approaches succeed in partially alleviating the problem by allowing the NEC to be satisfied at the throat, as shown in $f(R)$ gravity \cite{Lobo:2009ip}, conformal Weyl gravity \cite{Lobo:2008zu}, non-minimal curvature-matter coupling in $f(R)$ \cite{Garcia:2010xb, MontelongoGarcia:2010xd}, and modified teleparallel gravity \cite{Boehmer:2012uyw}. Thus, although the amount of exotic matter required can be significantly reduced, it still remains necessary to construct such solutions in these setups. Nevertheless, there are exceptional cases where both the NEC and the WEC are satisfied throughout the entire spacetime, such as in Einstein-Gauss-Bonnet gravity \cite{Mehdizadeh:2015jra}, third-order Lovelock gravity \cite{KordZangeneh:2015dks}, Ricci-based gravity \cite{Bouhmadi-Lopez:2018sto}, and hybrid metric-Palatini gravity \cite{Rosa:2018jwp}.

In this work, we focus on Symmetric Teleparallel Gravity (STG) in which the gravitational field is described through non-metricity rather than curvature or torsion \cite{Nester:1998mp}, and the underlying geometry is characterized by a flat, torsion-free affine connection. Consequently, the non-metricity scalar $Q$ \cite{Mol:2014ooa}, which quantifies how the length of a vector changes under parallel transport, plays the central role in the gravitational action \cite{BeltranJimenez:2017vop}. When the action is constructed linearly in $Q$, the resulting theory, known as the symmetric teleparallel equivalent of general relativity (STEGR), reproduces the dynamics of GR at the level of the field equations. Departing from this linear prescription leads to a broad class of modified gravity theories, among which $f(Q)$ gravity has attracted particular attention. In this framework, the gravitational Lagrangian is promoted to a general function of the non-metricity scalar, $f(Q)$, thereby extending STEGR beyond its direct equivalence with GR \cite{BeltranJimenez:2018vdo}. This construction provides a geometrically distinct route to modifying gravitational interactions while retaining second-order field equations for the metric in the standard formulation, in contrast to the higher-order equations typically encountered in generic $f(R)$ theories. A generalization of this type of theory can be achieved by extending $f(Q)$ to $f(Q,C)$, where $C$ denotes the difference between $Q$ and the Ricci scalar of GR \cite{De:2023xua, Capozziello:2023vne}. In STEGR, $C$ acts as a boundary term in the action and therefore does not contribute to the equations of motion. However, beyond the linear case, this term can give rise to new phenomenological effects \cite{Artola:2025fup, Ayuso:2026hqz}.

The versatility of $f(Q)$ gravity has motivated extensive research into its theoretical foundations and phenomenological applications, encompassing early- and late-time cosmology \cite{Frusciante:2021sio}, inflationary scenarios \cite{Capozziello:2022tvv, Capozziello:2024lsz},  gravitational-wave polarizations \cite{Dong:2025pyz}, and quantum cosmology \cite{Paliathanasis:2024hvl}. A substantial part of the literature has focused on cosmological applications, where appropriate choices of $f(Q)$ can account for the observed accelerated expansion of the Universe \cite{Lazkoz:2019sjl, Boiza:2025xpn} and have been investigated as potential means of alleviating tensions between cosmological observations \cite{Basilakos:2025olm, Boiza:2025xpn}. Furthermore, the framework allows for the reproduction of established cosmological solutions in GR, providing a basis for exploring deviations from the standard cosmological model \cite{Chakraborty:2025qlv, Lazkoz:2019sjl}. Various functional forms have been considered in this context. In particular, power-law corrections to the STEGR action were proposed as a possible explanation for dark energy \cite{BeltranJimenez:2019tme}, although models of this type without a cosmological constant have been found to be disfavoured by observational constraints \cite{Ayuso:2020dcu, Sahlu:2024pxk}. Similarly, Dvali--Gabadadze--Porrati (DGP)-inspired models and their extensions have been confronted with cosmological observations \cite{Ayuso:2021vtj}, while other power-law and exponential forms of $f(Q)$ have been shown to accommodate an accelerated expansion consistent with current data \cite{Dialektopoulos:2025ihc, Paliathanasis:2025hjw}. The relevance of the connection has also been investigated in recent works \cite{DAmbrosio:2021pnd, Shabani:2023xfn, Paliathanasis:2023pqp, Yang:2024tkw, Dimakis:2022wkj}, showing that different choices of connection can lead to markedly different cosmological dynamics, even for the same functional form of $f(Q)$ \cite{Dimakis:2022rkd, Ayuso:2025vkc}. 

Beyond cosmology, modified gravitational dynamics can also be investigated through the geometry of compact objects and nontrivial spacetimes. In this regard, black holes \cite{Bahamonde:2022zgj, Dimakis:2024fan, Zhang:2025tsv, Dehyadegari:2026vfd, Chen:2026lof} and relativistic stellar configurations \cite{Dimakis:2025jrl} have been studied within $f(Q)$ gravity, while wormhole solutions provide a particularly interesting setting to examine the interplay between the spacetime geometry, the gravitational dynamics, and the conditions required for the existence of traversable configurations. Nevertheless, these types of spacetime impose specific symmetries on the connection that constrain its form \cite{DAmbrosio:2021zpm}, an aspect that is often overlooked in studies of wormholes.

The paper is organized as follows. In Sec. \ref{sec:fq}, we introduce the symmetric teleparallel geometry and, in particular, $f(Q)$ gravity. In Sec. \ref{sec:wh}, we introduce the Ellis--Bronnikov wormhole spacetime and discuss its symmetries. These symmetries are then imposed on the affine connection in Sec. \ref{sec:connection}, where we find two possible families of connection components. We also derive the corresponding field equations for each family of connections. In Sec. \ref{sec:conn1sol}, we focus on the nontrivial solutions associated with Connection I, while the nontrivial solutions associated with Connection II are studied in Sec. \ref{sec:conn2sol}. Finally, in Sec. \ref{sec:conc}, we summarize our main results and present our conclusions.

\section{\texorpdfstring{$f(\mathbb Q)$ } <theory}\label{sec:fq}

The geometric structure of a spacetime is determined by two independent objects defined on the underlying manifold: the metric tensor $g_{\mu\nu}$, which encodes distances and angles, and the affine connection $\Gamma^{\alpha}{}_{\mu\nu}$, which prescribes how tensors are parallel-transported from one point to another.  In GR, these two objects are not independent: the connection is uniquely fixed to be the Levi-Civita connection $\left\{\substack{\alpha\\ \mu\nu}\right\}$, which is both metric-compatible and torsion-free.  However, this identification is a choice rather than a necessity, and relaxing it opens the door to a broad landscape of gravitational theories~\cite{Hehl:1994ue, BeltranJimenez:2019esp}.

A general affine connection can be decomposed as~\cite{Jarv:2018bgs, BeltranJimenez:2019esp}
\bea
\Gamma^{\alpha}{}_{\mu\nu}=\left\{\substack{\alpha\\ \mu\nu}\right\}+K^{\alpha}{}_{\mu\nu}+L^{\alpha}{}_{\mu\nu}\,,
\eea
where
\bea
K^{\alpha}{}_{\mu\nu}=\frac{1}{2}T^{\alpha}{}_{\mu\nu}+T_{(\mu}{}^{\alpha}{}_{\nu)}
\eea
is the contortion tensor, built from the torsion
\begin{equation}
    T^{\alpha}{}_{\mu\nu}\equiv\Gamma^\alpha{}_{\mu\nu}-\Gamma^\alpha{}_{\nu\mu}\,,
\end{equation}
and
\bea
L^{\alpha}{}_{\mu\nu}=\frac{1}{2}Q^{\alpha}{}_{\mu\nu}-Q_{(\mu}{}^{\alpha}{}_{\nu)}
\eea
is the disformation tensor, built from the non-metricity tensor
\begin{equation}
    Q_{\alpha\mu\nu}\equiv\nabla_\alpha g_{\mu\nu}\,.
\end{equation}
The non-metricity tensor quantifies the failure of the metric to be covariantly constant, or equivalently, measures how the length of a vector changes under parallel transport~\cite{Mol:2014ooa}.

From the non-metricity tensor one can define two independent traces,
\bea
Q_\alpha = g^{\mu\nu}Q_{\alpha\mu\nu}\,,\qquad \tilde{Q}_\alpha = g^{\mu\nu}Q_{\mu\alpha\nu}\,,
\eea
which, together with $Q_{\alpha\mu\nu}$ itself, can be used to construct the non-metricity scalar.  In general, the most general parity-even scalar quadratic in the non-metricity tensor involves five free parameters~\cite{BeltranJimenez:2017tkd, BeltranJimenez:2019esp}.  In this work, we adopt the standard choice\footnote{An equivalent definition, commonly found in the literature, differs by an overall sign and reads $Q = -Q_{\alpha\mu\nu}Q^{\alpha\mu\nu}/4 + Q_{\alpha\mu\nu}Q^{\mu\alpha\nu}/2 + Q_\alpha Q^\alpha/4 - Q_\alpha \tilde{Q}^\alpha/2$.  This convention does not affect the physics.}
\bea
Q = \frac{1}{4}Q_{\alpha\mu\nu}Q^{\alpha\mu\nu} - \frac{1}{2}Q_{\alpha\mu\nu}Q^{\mu\alpha\nu} - \frac{1}{4}Q_\alpha Q^\alpha + \frac{1}{2}Q_\alpha \tilde{Q}^\alpha\,.\label{Qscalar}
\eea
This particular combination is singled out by the requirement that, in the Symmetric Teleparallel Equivalent of General Relativity (STEGR)~\cite{Nester:1998mp, BeltranJimenez:2017tkd}, the non-metricity scalar reproduces the Einstein--Hilbert dynamics.  More precisely, by choosing a connection that is both flat and torsion-free, one can show that the Levi-Civita Ricci scalar and the non-metricity scalar are related by~\cite{BeltranJimenez:2017tkd, Jarv:2018bgs}
\bea
R(\{\})=-Q - \nabla^{\{\}}_\alpha(Q^\alpha-\tilde Q^\alpha)\,,
\label{RQ}
\eea
where $\{\}$ denotes the Levi-Civita connection.  Since the last term is a total divergence, STEGR and GR yield identical field equations, despite being built from entirely different geometric quantities.

It is useful to introduce the non-metricity conjugate, or superpotential tensor~\cite{BeltranJimenez:2019tme},
\bea
P^{\alpha}{}_{\mu\nu} &=& \frac{1}{4}\Big[-Q^{\alpha}{}_{\mu\nu} + 2 Q_{(\mu}{}^{\alpha}{}_{\nu)} + Q^{\alpha}g_{\mu\nu}\nonumber\\
&& - \tilde{Q}^{\alpha}g_{\mu\nu} - \delta^{\alpha}{}_{(\mu}Q_{\nu)}\Big]\,,\label{superpotential}
\eea
in terms of which the non-metricity scalar takes the compact form $Q = -Q_{\alpha\mu\nu}P^{\alpha\mu\nu}$.

The equivalence between GR and STEGR at the level of the equations of motion motivates a natural generalization: promoting the linear dependence on $Q$ to an arbitrary function $f(Q)$, in direct analogy with the passage from GR to $f(R)$ gravity.  The action for $f(Q)$ gravity reads~\cite{BeltranJimenez:2019tme}
\begin{equation}
    S=\int{d^4 x\sqrt{-g}\left[-\frac{1}{2}f(Q)+\mathcal{L}_{M}\right]}\,,\label{fQaction}
\end{equation}
where STEGR is recovered for $f(Q) = Q/\kappa$, with $\kappa = 8\pi G$.  It is important to emphasize that, although Eq.~\eqref{RQ} implies the dynamical equivalence of STEGR with GR, this equivalence does \emph{not} extend to nonlinear generalizations: the boundary term $\nabla^{\{\}}_\alpha(Q^\alpha-\tilde Q^\alpha)$ in Eq.~\eqref{RQ} is harmless in the linear case but contributes nontrivially once $f$ is a nonlinear function.  Consequently, $f(Q)$ gravity and $f(R)$ gravity are genuinely distinct theories with different dynamics and different solution spaces~\cite{BeltranJimenez:2019tme, Zhao:2021zab}.

Variation of the action~\eqref{fQaction} with respect to the metric yields the gravitational field equations~\cite{BeltranJimenez:2019tme}: 
\bea
\mathcal{M}_{\mu\nu}&&\equiv\frac{2}{\sqrt{-g}}\nabla_\alpha\left(\sqrt{-g}f_QP^\alpha{}_{\mu\nu}\right)+\frac{1}{2}g_{\mu\nu}f\nonumber\\
&&+f_Q\left(P_{\mu\alpha\beta}Q_\nu{}^{\alpha\beta}-2Q_{\alpha\beta\mu}P^{\alpha\beta}{}_{\nu}\right)=T_{\mu\nu}\,,\label{met.eq}
\eea
where $f_Q\equiv d f/dQ$.  Variation with respect to the connection, under the assumptions of vanishing torsion, vanishing curvature, and absence of hypermomentum, gives~\cite{BeltranJimenez:2019tme}
\bea
\mathcal{C}_\alpha\equiv\nabla_\mu\nabla_\nu\left(\sqrt{-g}f_QP^{\mu\nu}{}_{\alpha}\right)=0\,.
\label{eq:con}
\eea
The system~\eqref{met.eq}--\eqref{eq:con} constitutes the complete set of equations of motion.  A distinctive feature of $f(Q)$ gravity, compared to metric $f(R)$ theories, is the presence of the connection field equation~\eqref{eq:con}, which constrains the affine structure independently of the metric.  This additional equation plays a central role in the analysis that follows, as it restricts the allowed connection configurations for a given spacetime geometry.

\section{Wormhole of Ellis-Bronnikov}\label{sec:wh}

Among the plethora of different wormhole geometries, the Ellis--Bronnikov wormhole is one of the simplest and most well-studied traversable wormhole solutions in gravitational physics.  It was discovered independently by Ellis~\cite{Ellis:1973yv} and Bronnikov~\cite{Bronnikov:1973fh} in 1973, and later revisited in the seminal work of Morris and Thorne~\cite{Morris:1988cz}, which laid the foundations for the modern study of traversable wormholes.
 
In its original form, the line element is written in terms of the proper radial coordinate $\rho\in(-\infty,+\infty)$ as
\bea
ds^2=-dt^2+d\rho^2+(\rho^2+a^2)(d\theta^2+\sin^2\theta\, d\phi^2)\,,\label{metricrho}
\eea
where $a>0$ is a constant.  The coordinate $\rho$ covers the entire spacetime: it runs from $-\infty$ on one asymptotically flat side, through $\rho=0$, and out to $+\infty$ on the other.  The area of the two-spheres of constant $t$ and $\rho$ is $4\pi(\rho^2+a^2)$, which attains its minimum value $4\pi a^2$ at $\rho=0$.  This minimal-area sphere defines the wormhole throat, and the parameter $a$ is therefore its areal radius.

Several features of this geometry are worth noting.  The metric is regular everywhere, with no horizons or singularities.  The redshift function vanishes identically ($g_{tt}=-1$), ensuring that a static observer at any point experiences no gravitational redshift relative to spatial infinity.  This is what makes the wormhole traversable: an observer can pass through the throat in finite proper time and without encountering infinite tidal forces~\cite{Morris:1988cz, Visser:1995cc}.

It is convenient to introduce an areal radial coordinate $r$ defined by $r^2=\rho^2+a^2$, which ranges from the throat at $r=a$ to spatial infinity at $r\to\infty$ on each side.  In terms of $r$, the metric~\eqref{metricrho} becomes
\bea
ds^2=-dt^2+\frac{r^2}{r^2-a^2}\,dr^2+r^2(d\theta^2+\sin^2\theta\, d\phi^2)\,.\label{metric}
\eea
This form can be directly compared with the general Morris--Thorne ansatz for a static, spherically symmetric wormhole~\cite{Morris:1988cz}:
\bea
ds^2=-e^{2\Phi(r)}dt^2+\frac{dr^2}{1-b(r)/r}+r^2\,d\Omega^2\,,\label{morristhorne}
\eea
where $\Phi(r)$ is the redshift function and $b(r)$ is the shape function.  For the Ellis--Bronnikov geometry one reads off
\bea
\Phi(r)=0\,,\qquad b(r)=\frac{a^2}{r}\,.\label{shapefunc}
\eea
The throat is located where $b(r_0)=r_0$, which gives $r_0=a$, consistent with the discussion above.

A well-defined wormhole throat must satisfy the \emph{flaring-out condition}~\cite{Morris:1988cz, Hochberg:1997wp}, which demands that the embedding surface flares outward at the throat.  In terms of the shape function, this condition reads
\bea
b'(r_0)<1\,,\label{flaringout}
\eea
at the throat $r_0$, where a prime denotes differentiation with respect to $r$.  For the Ellis--Bronnikov shape function~\eqref{shapefunc}, one finds $b'(r)=-a^2/r^2$, so that $b'(a)=-1$, which comfortably satisfies the bound~\eqref{flaringout}.  However, it is precisely this condition that, in the framework of GR, leads to a violation of the NEC~\cite{Morris:1988cz, Hochberg:1997wp}.  Indeed, combining the flaring-out condition with the Einstein field equations at the throat yields $\rho+p_r\big|_{r_0}<0$, where $\rho$ is the energy density and $p_r$ the radial pressure, signaling the presence of exotic matter~\cite{Visser:1995cc}.

Within GR, the Ellis--Bronnikov wormhole is supported by a massless phantom scalar field, i.e.\ a scalar field with a reversed (negative) kinetic term~\cite{Ellis:1973yv, Bronnikov:1973fh}.  While such phantom fields are commonly invoked in wormhole physics, they are problematic from a theoretical standpoint: they lead to an energy spectrum unbounded from below and are typically associated with vacuum instabilities~\cite{Cline:2003gs, Carroll:2003st}.  This provides a strong motivation for investigating whether the same wormhole geometry can be sustained without exotic matter in the context of modified gravity theories, where the additional gravitational degrees of freedom may effectively play the role of the phantom field.  This is the strategy we pursue in the present work within $f(Q)$ gravity.

\begin{figure}
    \centering
    \includegraphics[width=1\linewidth]{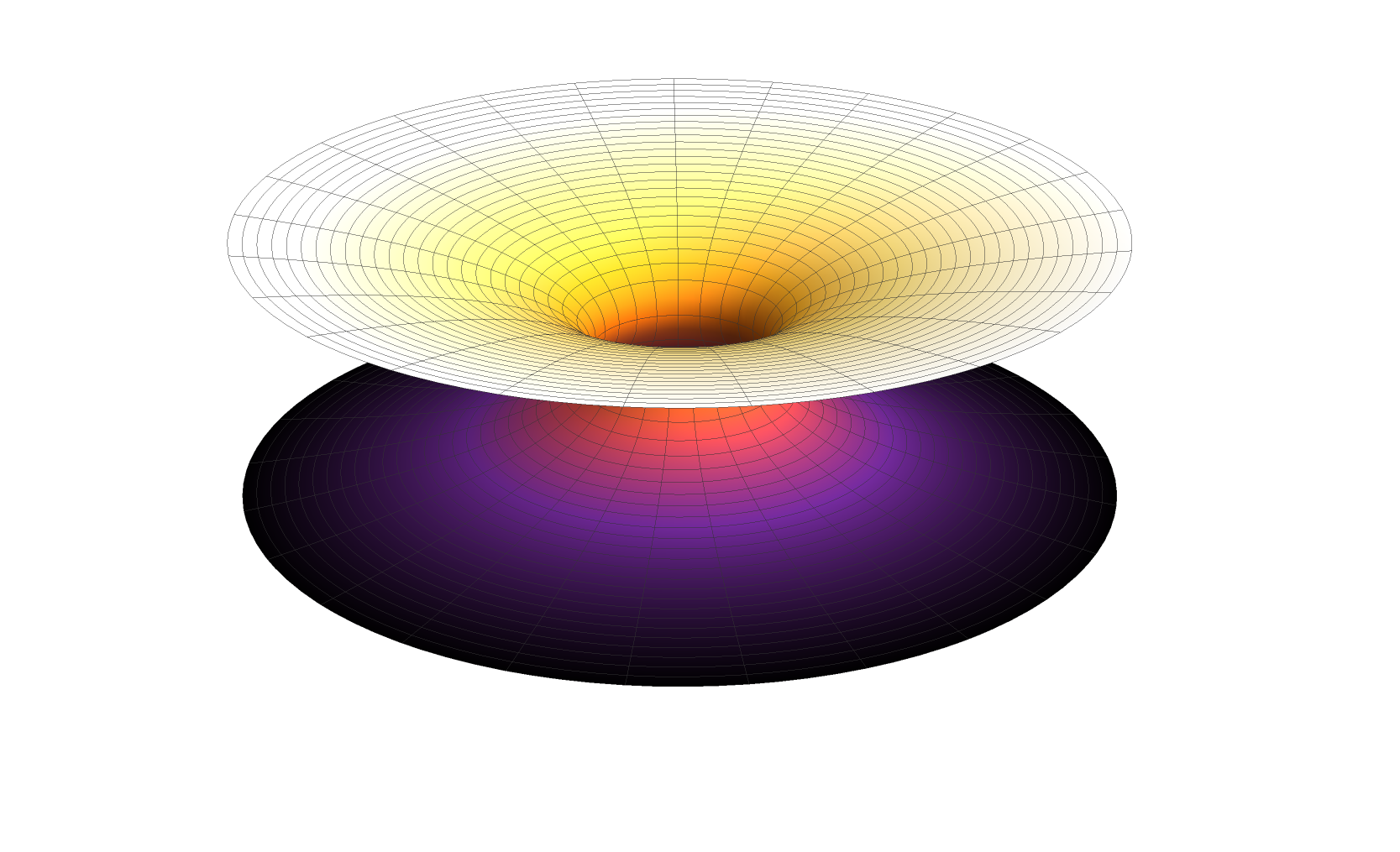}
    \caption{Embedding diagram of the Ellis--Bronnikov wormhole.}
    \label{fig:embedding}
\end{figure}

To illustrate the geometry of the Ellis--Bronnikov wormhole, Fig.~\ref{fig:embedding} shows its embedding diagram in a three-dimensional Euclidean space. The two asymptotically flat regions are connected through the wormhole throat, located at the minimum of the embedding surface. The characteristic flaring-out of the surface at the throat provides a direct geometrical visualization of the flaring-out condition discussed above.

\subsection{Spacetime symmetries}
 
The metric~\eqref{metric} is static and spherically symmetric.  Its isometry group is generated by the following Killing vector fields~\cite{Heisenberg:2023lru}: The generator of time translations:
\bea
\mathcal{T}=\partial_t\,,\label{killingT}
\eea
and the generators of spatial rotations:
\bea
\mathcal{R}_x&=&\sin\phi\,\partial_\theta+\frac{\cos\phi}{\tan\theta}\,\partial_\phi\,,\nonumber\\
\mathcal{R}_y&=&-\cos\phi\,\partial_\theta+\frac{\sin\phi}{\tan\theta}\,\partial_\phi\,,\nonumber\\
\mathcal{R}_z&=&-\partial_\phi\,.\label{killingR}
\eea
These satisfy the Killing equation
\bea
\mathcal{L}_\xi g_{\mu\nu}=0\,,\label{liemetric}
\eea
for $\xi\in\{\mathcal{T},\mathcal{R}_x,\mathcal{R}_y,\mathcal{R}_z\}$.  In the following section, we will require the affine connection to respect the same symmetries, a condition that plays a central role in constraining its components.

\subsection{Matter content}
 
We assume that the matter supporting the wormhole geometry is described by an anisotropic perfect fluid, with the stress-energy tensor given by
\bea
T^{\mu}{}_{\nu}=\text{diag}(-\rho,\, p_r,\, p_t,\, p_t)\,,
\label{eq:Tmunu}
\eea
where $\rho$ is the energy density, $p_r$ is the radial pressure (along the $r$-direction), and $p_t$ is the tangential pressure (along the angular directions).  Anisotropic matter distributions arise naturally in the study of wormholes, since the distinct roles played by the radial and angular directions at the throat generically require $p_r\neq p_t$~\cite{Morris:1988cz, Lobo:2005us}.  In the context of $f(Q)$ gravity, the effective stress-energy tensor supporting the spacetime metric receives additional contributions from the non-metricity sector, and it is the interplay between these geometric terms and the matter content that determines whether the energy conditions can be satisfied.

\section{Affine connection and field equations for the Ellis--Bronnikov wormhole}\label{sec:connection}

A defining feature of $f(Q)$ gravity is that the affine connection is treated as an independent dynamical variable, subject to the constraints of vanishing torsion and vanishing curvature. Unlike in GR, where the connection is uniquely determined by the metric through the Levi-Civita prescription, the symmetric teleparallel framework admits a family of flat, torsion-free connections for any given metric. Different members of this family are physically equivalent in STEGR, where they are related by the full affine symmetry group of the theory. However, this equivalence is generically broken in $f(Q)$ gravity: the nonlinearity of $f$ reduces the symmetry group, and distinct connections lead to genuinely different field equations and solutions~\cite{BeltranJimenez:2019tme, Zhao:2021zab, DAmbrosio:2021zpm}. Identifying the allowed connections for a given spacetime geometry is therefore a necessary first step in any $f(Q)$ analysis.

In practice, a common approach in the literature is to work in the so-called \emph{coincident gauge}, in which all connection components vanish: $\Gamma^\alpha{}_{\mu\nu}=0$. Since the connection is flat, such a coordinate system always exists by virtue of a theorem in affine geometry. However, the coordinate transformation required to reach the coincident gauge generically changes the form of the metric~\cite{BeltranJimenez:2019esp}. In particular, for the Ellis--Bronnikov metric written in the areal coordinates of Eq.~\eqref{metric}, the coincident gauge is not compatible with the standard form of the line element. Consequently, we must work with a non-vanishing connection, and we determine its components systematically by imposing the defining constraints of the symmetric teleparallel framework together with the spacetime symmetries, following the approach of Ref.~\cite{DAmbrosio:2021zpm}.

\subsection{Systematic determination of the connection}

We proceed in three stages, progressively reducing the number of independent connection components.

\paragraph{Step 1: Vanishing torsion.}
We impose the symmetry of the connection in its lower indices:
\bea
T^{\alpha}{}_{\mu\nu}\equiv 2\,\Gamma^\alpha{}_{[\mu\nu]}\overset{!}{=}0\,.
\eea
In four spacetime dimensions, a general connection has $4^3=64$ independent components. The torsion-free condition provides $4\times\binom{4}{2}=24$ independent constraints, reducing the number of free components to $40$.

\paragraph{Step 2: Spacetime symmetries.}
We require the connection to be invariant under the same isometries as the metric. Concretely, for each Killing vector field $\xi$ of the metric~\eqref{metric}, we demand
\bea
\mathcal{L}_\xi \Gamma^{\alpha}{}_{\mu\nu}\overset{!}{=}0\,,
\label{lieconnection}
\eea
where the Lie derivative of the connection is given by~\cite{Dimakis:2023uib}
\bea
\mathcal{L}_\xi \Gamma^{\alpha}{}_{\mu\nu}&=&\xi^k\partial_k \Gamma^{\alpha}{}_{\mu\nu}-\Gamma^{k}{}_{\mu\nu}\partial_k \xi^\alpha+\Gamma^{\alpha}{}_{k\nu}\partial_\mu \xi^k\nonumber\\
&&+\Gamma^{\alpha}{}_{\mu k}\partial_\nu \xi^k+\partial_\mu\partial_\nu \xi^\alpha\,.
\label{liederivative}
\eea
This expression differs from the Lie derivative of a tensor by the last term, reflecting the inhomogeneous transformation law of the connection. Imposing Eq.~\eqref{lieconnection} for each of the four Killing vectors~\eqref{killingT}--\eqref{killingR} proceeds as follows. The time-translation generator $\mathcal{T}$ requires all connection components to be independent of $t$, while the axial rotation generator $\mathcal{R}_z$ requires independence of $\phi$. The remaining generators $\mathcal{R}_x$ and $\mathcal{R}_y$ impose additional algebraic and differential relations among the components. In total, these symmetry conditions fix $28$ of the $40$ torsion-free components, leaving $12$ undetermined.

\paragraph{Step 3: Vanishing curvature.}
Finally, we impose the flatness condition:
\bea
R^\alpha{}_{\beta\mu\nu}\overset{!}{=}0\,.
\eea
This yields $24$ non-trivial equations---a mixture of algebraic relations and first-order ordinary differential equations---for the $12$ remaining components. The system admits two distinct families of solutions, which we present below\footnote{These two families exhaust the space of flat, torsion-free connections compatible with the static and spherically symmetric isometries of the metric~\eqref{metric}.}.

\subsection{Connection I}\label{sec:conn1}

The first family of solutions is parametrized by one constant $c$ and two free functions $\alpha(r)$, $\xi(r)$. All connection components not listed below vanish, and the torsion-free condition $\Gamma^\alpha{}_{\mu\nu}=\Gamma^\alpha{}_{\nu\mu}$ is understood throughout. The independent nonzero components are:
\bea
&&\Gamma^t{}_{tt}=c\,,\quad \Gamma^t{}_{tr}=\alpha\,,\quad \Gamma^t{}_{rr}=\gamma\,,\nonumber\\
&&\Gamma^t{}_{\theta\theta}=-\frac{1}{c}\,,\quad \Gamma^t{}_{\phi\phi}=-\frac{\sin^2\theta}{c}\,,\nonumber\\
&&\Gamma^r{}_{rr}=\xi\,,\nonumber\\
&&\Gamma^\theta{}_{t\theta}=c\,,\quad \Gamma^\theta{}_{r\theta}=\alpha\,,\quad \Gamma^\theta{}_{\phi\phi}=-\cos\theta\sin\theta\,,\nonumber\\
&&\Gamma^\phi{}_{t\phi}=c\,,\quad \Gamma^\phi{}_{r\phi}=\alpha\,,\quad \Gamma^\phi{}_{\theta\phi}=\cot\theta\,,
\label{conn1components}
\eea
where $\gamma(r)$ is determined by the flatness condition:
\bea
\gamma(r)=\frac{\alpha'(r)-\alpha(r)\left[\xi(r)-\alpha(r)\right]}{c}\,.
\eea
The constant $c$ must be nonzero, as it appears in several denominators. The structure of this connection is notable for its simplicity: the angular components $\Gamma^\theta{}_{\phi\phi}$ and $\Gamma^\phi{}_{\theta\phi}$ coincide with those of the Levi-Civita connection on the two-sphere, while the temporal and radial sectors share the same functions $c$ and $\alpha(r)$. Having determined the components of the affine connection, we can now derive the corresponding field equations and the non-metricity scalar.

\paragraph{Off-Diagonal metric equations.} The diagonal form of the stress-energy tensor~\eqref{eq:Tmunu} requires all off-diagonal components of the left-hand side of Eq.~\eqref{met.eq} to vanish. For Connection~I, this condition is automatically satisfied for every component except $\mathcal{M}^{t}{}_{r}$, which yields
\bea
\mathcal{M}^{t}{}_{r}\equiv-\frac{3}{2}\, c\, Q'(r)\,f_{QQ}=0\,.
\label{eq:mtrconn1}
\eea
Since $c\neq0$, there are two possible cases. The first is $f_{QQ}=0$ (apart from the special case $Q'=0$), which gives $f(Q)=a_0+a_1 Q,$ with $a_0$ and $a_1$ constants. This choice reduces the theory to GR with a cosmological constant. The second possibility is $Q'(r)=0$.

The diagonal components of the metric field equation~\eqref{met.eq}, together with the connection field equation~\eqref{eq:con}, constitute the dynamical system that must be solved. 
\paragraph{Diagonal metric equations.} These equations relate the geometry to the matter content and are given by:
\bea
\mathcal{M}^{t}{}_{t}&\equiv& \frac{1}{2}f-f_Q\left(\frac{Q}{2}-\frac{a^2}{r^4}\right)\nonumber\\
&&+\frac{(a^2-r^2)(3 r \alpha-4)}{2 r^3}Q'(r)f_{QQ}=-\rho
\,,\label{eq:mtteq-1-gen}
\eea
\bea
\mathcal{M}^{r}{}_{r}&\equiv&\frac{1}{2}f+f_Q\left(-\frac{Q}{2}-\frac{a^2}{r^4}\right)\nonumber\\
&&+\frac{3(r^2-a^2)\alpha}{2 r^2}Q'(r)f_{QQ}=p_r
\,,\label{eq:mrreq-1-gen}
\eea
\bea
\mathcal{M}^{\theta}{}_{\theta}&\equiv&\frac{1}{2}f+f_Q\left(-\frac{Q}{2}+\frac{a^2}{r^4}\right)\nonumber\\
&&+\frac{(r^2-a^2)(3 r \alpha-2)}{2 r^3}Q'(r)f_{QQ}=p_t
\,.\label{eq:mthetheeq-1-gen}
\eea

\paragraph{Connection field equations.}
The connection equation~\eqref{eq:con} comprises four components. For the Ellis--Bronnikov metric with Connection~I, the angular components, $\mathcal{C}_\theta=0$ and $\mathcal{C}_\phi=0$, are identically satisfied. The remaining nontrivial equations are $\mathcal{C}_t$ and $\mathcal{C}_r$:
\begin{align}
    \mathcal{C}_t \equiv&\frac{3 c }{4}r\sqrt{\left(r^2-a^2\right)\sin^2\theta}\nonumber\\
    &\times\left[\left(\frac{2r^2-a^2}{r(r^2-a^2)}Q'(r)+Q''(r)\right)f_{QQ}+Q'^2(r)f_{QQQ}\right]\,,\label{eq:Ct_I}
\end{align}
\begin{align}
    \mathcal{C}_r \equiv&\frac{3 \alpha }{4}r\sqrt{\left(r^2-a^2\right)\sin^2\theta}\nonumber\\
    &\times\left[\left(\frac{2r^2-a^2}{r(r^2-a^2)}Q'(r)+Q''(r)\right)f_{QQ}+Q'^2(r)f_{QQQ}\right]\,,\label{eq:Cr_I}
\end{align}
with $\mathcal{C}_t=\mathcal{C}_{r}\,(c/\alpha)$, so only one of the two is independent.

\paragraph{Non-metricity scalar.}
The non-metricity scalar $Q$ for the Ellis--Bronnikov metric with Connection~I evaluates to:
\begin{equation}
    Q(r)=\frac{\left(a^2-2 r^2\right) (2-3 r \alpha ) -3 r^2
   \left(a^2-r^2\right) \alpha '}{r^4}\,.\label{eq:Q_conI}
\end{equation}

It should be stressed that neither $\xi(r)$ nor $\gamma(r)$ enters the field equations or the non-metricity scalar \eqref{eq:Q_conI}. This follows from the fact that $Q(r)$ depends only on $\alpha(r)$ and its derivative: the component $\Gamma ^r{}_{rr} = \xi(r)$ contracts to zero against the diagonal Ellis-Bronnikov metric in the non-metricity tensor, while $\gamma(r)$ is algebraically determined by $\alpha$ and $\xi$ via the flatness condition. As a result, $\xi(r)$ remains a genuinely free but physically inert function; any choice is compatible with the dynamics.
\paragraph{Summary: the reduced system.}
To summarize, the complete system for Connection~I reads
\bea
&\mathcal{M}^{t}{}_{r}=0\,,\quad\mathcal{M}^{t}{}_{t}=-\rho\,,\quad \mathcal{M}^{r}{}_{r}=p_r\,,&\nonumber\\ &\mathcal{M}^{\theta}{}_{\theta}=p_t\,,\quad \mathcal{C}_r=0\,,&
\label{eq:reducedsystemI}
\eea
since $\mathcal{C}_r=0$ automatically satisfies  $\mathcal{C}_t=0$.

\subsection{Connection II}\label{sec:conn2}

The second family is parametrized by two constants $c$ and $k$, and two free functions $\alpha(r)$ and $\beta(r)$. We introduce the shorthand $d\equiv 2c-k$ for compactness. The independent nonzero components are:{\footnote{The trivial spherical connection widely considered in the literature, e.g., Refs.~\cite{Zhao:2021zab,Lin:2021uqa} and the references therein, belongs to this class by taking $c=k=\alpha=0$ and $\beta=-r$. In this case, the connection degree of freedom is completely turned off, hence plays no role in the system \cite{DAmbrosio:2021zpm}.}
\bea
&&\Gamma^t{}_{tt}=-c+k-cd\,\alpha\,,\nonumber\\
&&\Gamma^t{}_{tr}=\frac{d\,\alpha(1+c\,\alpha)}{\beta}\,,\quad \Gamma^t{}_{rr}=\gamma\,,\nonumber\\
&&\Gamma^t{}_{\theta\theta}=\alpha\,,\quad \Gamma^t{}_{\phi\phi}=\alpha\,\sin^2\theta\,,\nonumber\\[6pt]
&&\Gamma^r{}_{tt}=-c\,d\,\beta\,,\quad \Gamma^r{}_{tr}=c+c\,d\,\alpha\,,\nonumber\\
&&\Gamma^r{}_{rr}=\xi\,,\nonumber\\
&&\Gamma^r{}_{\theta\theta}=\beta\,,\quad \Gamma^r{}_{\phi\phi}=\beta\,\sin^2\theta\,,\nonumber\\[6pt]
&&\Gamma^\theta{}_{t\theta}=c\,,\quad \Gamma^\theta{}_{r\theta}=-\frac{1+c\,\alpha}{\beta}\,,\nonumber\\
&&\Gamma^\theta{}_{\phi\phi}=-\cos\theta\sin\theta\,,\nonumber\\[6pt]
&&\Gamma^\phi{}_{t\phi}=c\,,\quad \Gamma^\phi{}_{r\phi}=-\frac{1+c\,\alpha}{\beta}\,,\nonumber\\
&&\Gamma^\phi{}_{\theta\phi}=\cot\theta\,,
\label{conn2components}
\eea
where the functions $\gamma(r)$ and $\xi(r)$ are fixed by the flatness condition:
\bea
\gamma(r)=\frac{1}{\beta}\left[-\frac{\alpha}{\beta}\left(1+c\alpha\right)\left(1+2c\alpha-k\alpha\right)-\alpha'\right]\,,
\label{gamma2}
\eea
\bea
\xi(r)&=&\frac{1}{\beta}\left[-1-c\,\alpha\left(2+(2c-k)\alpha\right)-\beta'\right].
\label{xi2}
\eea
This family has a considerably richer structure than Connection~I. Most notably, the components $\Gamma^\theta{}_{r\theta}$ and $\Gamma^\phi{}_{r\phi}$ involve the ratio $(1+c\,\alpha)/\beta$, which will play a crucial role in the regularity analysis at the wormhole throat. Furthermore, the $(t,r)$-sector now involves both constants $c$ and $k$, as well as both free functions $\alpha(r)$ and $\beta(r)$, providing substantially more freedom to accommodate nontrivial $f(Q)$ solutions.

It is worth emphasizing that the constants $c$ and $k$ are not gauge artifacts. In STEGR, different values of these parameters are related by the affine symmetry group and lead to equivalent physics. In $f(Q)$ gravity, however, the nonlinearity of $f$ breaks this equivalence: the field equations depend explicitly on $c$ and $k$ through the non-metricity scalar and its derivatives, and different values generically yield distinct solution spaces~\cite{DAmbrosio:2021zpm, Heisenberg:2023lru}. As we will show, the physically relevant quantity is ultimately the dimensionless combination $ak$, where $a$ is the throat radius.

\paragraph{Off-diagonal metric equations.} Connection~II proves far more fruitful. Again, the only non-vanishing off-diagonal metric field equation is $\mathcal{M}^{t}{}_{r}$, which now reads
\bea
\mathcal{M}^{t}{}_{r}\equiv -\frac{1}{2}\left[k-2c\,d\,\alpha\right]Q'(r)\,f_{QQ}=0\,,
\label{eq:mtrconn2}
\eea
where $d\equiv 2c-k$ as before. Unlike the Connection~I case, this equation can be satisfied without forcing $f_{QQ}=0$ or $Q'=0$, i.e., when $k-2cd\alpha=0$.

\paragraph{Diagonal metric equations.}
Before writing the diagonal metric equations explicitly, it is convenient to introduce the shorthand
\bea
\psi\equiv(a^2-r^2)(1+c\,\alpha)(2-d\,\alpha)\,,
\label{psidef}
\eea
which collects the dependence on the connection function $\alpha$ and the metric parameter $a$. As we will show in Sec.~\ref{sec:conn2sol}, regularity of the connection at the throat requires $1+c\,\alpha=0$, which immediately sets $\psi=0$ and substantially simplifies the system. We nevertheless present the equations in their general form, both for completeness and because the structure of the $\psi$-dependent terms clarifies the role played by the connection degrees of freedom.

The symmetry $\mathcal{M}^{\theta}{}_{\theta}=\mathcal{M}^{\phi}{}_{\phi}$ holds identically, so the three independent diagonal equations are:
\bea
\mathcal{M}^{t}{}_{t}&\equiv& \frac{1}{2}f-f_Q\left(\frac{Q}{2}-\frac{a^2}{r^4}\right)\nonumber\\
&&-\frac{r\,\psi+4(a^2-r^2)\beta+r(-2+c\,d\,r^2)\beta^2}{2r^3\beta}\nonumber\\
&&\times\, f_{QQ}\,Q'(r)=-\rho\,,\label{mtteq}
\eea
\bea
\mathcal{M}^{r}{}_{r}&\equiv&\frac{1}{2}f+f_Q\left(-\frac{Q}{2}-\frac{a^2}{r^4}\right)\nonumber\\
&&+\frac{\psi+(2+c\,d\,r^2)\beta^2}{2r^2\beta}\,f_{QQ}\,Q'(r)=p_r\,,\label{mrreq}
\eea
\bea
\mathcal{M}^{\theta}{}_{\theta}&\equiv&\frac{1}{2}f+f_Q\left(-\frac{Q}{2}+\frac{a^2}{r^4}\right)\nonumber\\
&&-\frac{r\,\psi+2(a^2-r^2)\beta-c\,d\,r^3\beta^2}{2r^3\beta}\nonumber\\
&&\times\, f_{QQ}\,Q'(r)=p_t\,.\label{mthetheeq}
\eea
Each equation contains a ``background'' part---the first line, involving $f$ and $f_Q$---that depends only on $Q$ and the metric, and a ``correction'' part---the remaining lines, proportional to $f_{QQ}\,Q'(r)$---that encodes the dynamical coupling to the connection through $\beta$ and $\psi$.

\paragraph{Connection field equations.}
Similar to Connection~I, $\mathcal{C}_\theta=0$ and $\mathcal{C}_\phi=0$ are automatically satisfied for the Ellis--Bronnikov metric with Connection~II. The two remaining nontrivial components are $\mathcal{C}_t$ and $\mathcal{C}_r$:

\begin{widetext}
\bea
\mathcal{C}_t&\equiv& \frac{|\sin\theta|}{4\sqrt{r^2-a^2}}\Big(\left[(a^2-2r^2)(2 c\, d\, \alpha-k)+2 c\,d\,(a^2-r^2)r\alpha'\right]Q'f_{QQ}\nonumber\\
&&+r(a^2-r^2)(2 c \,d\,\alpha-k)\frac{d}{dr}\Big[Q'(r)\,f_{QQ}\Big]\Big)=0\,.
\eea

\begin{align}
\mathcal{C}_r \equiv& -\frac{|\sin\theta|}{4\beta(r^2-a^2)^{3/2}}\left(
r(a^2-r^2)\Big[\psi+\beta^2(c\,d\,r^2+2)\Big]\frac{d}{dr}\Big[Q'(r)\,f_{QQ}\Big] \right.\nonumber\\
&\left.+\Big\{(a^2-r^2)\Big[(2-d\,\alpha)(a^2-2r^2)(c\,\alpha+1)+2\beta\,r\,\beta'(c\,d\,r^2+2)\Big]+\beta^2\Big[a^2(3c\,d\,r^2+2)-2c\,d\,r^4\Big]\Big\}Q'(r)\,f_{QQ}\right)=0\,.\label{b6}
\end{align}
\end{widetext}

The radial component $\mathcal{C}_r$ is considerably more involved than $\mathcal{C}_t$ and where we have used the abbreviation $\psi$ defined in Eq.~\eqref{psidef}. This equation is a second-order ODE for $\beta(r)$ (through $Q'(r)$, which itself depends on $\beta$ and $\beta'$), and its solution determines the connection profile.

Before continuing, we would like to mention that if one assumes $2cd\alpha-k=0$ to fulfill $\mathcal{M}^t{}_{r}=0$, substituting it into the expression above reduces $\mathcal{C}_t$ to
\bea
\mathcal{C}_t&\equiv&- \frac{|\sin\theta|\,c^2\;r\sqrt{r^2-a^2}\; \alpha'\,Q'\,f_{QQ}}{(1+2c\,\alpha)}=0\,.\label{c_t}
\eea
In this case, we have $\alpha'=0$, and the equation $\mathcal{C}_t=0$ is trivially satisfied, placing no further restriction on the system. This particular choice simplifies the system significantly and is helpful later to find nontrivial solutions in the case where $Q'(r)f_{QQ}$ does not vanish (see Sec.~\ref{sec:conn2sol}).

\paragraph{Non-metricity scalar.}
For completeness, we give the explicit expression for the non-metricity scalar $Q$ for  the Ellis--Bronnikov metric with Connection~II:
\begin{widetext}
\bea
Q(r)&=&\frac{1}{r^4\beta^2}\Bigg\{-\beta\left[(a^2-2r^2)\Big[r(1+c\,\alpha)(-2+d\,\alpha)-2\beta\Big]+\frac{2c\,d\,r^4-a^2(2+3c\,d\,r^2)}{r^2-a^2}\,r\,\beta^2\right]\nonumber\\
&&+r^2\left[(a^2-r^2)(k-2c\,d\,\alpha)\,\beta\,\alpha'+\Big[(r^2-a^2)(1+c\,\alpha)(2-d\,\alpha)-(2+c\,d\,r^2)\beta^2\Big]\beta'\right]\Bigg\}\,.\label{b7q}
\eea
\end{widetext}
This expression depends on both connection functions $\alpha(r)$ and $\beta(r)$, as well as on the constants $c$ and $k$.

\paragraph{Summary: the reduced system.}
To summarize, the complete system for Connection~II to be solved reads
\bea
&\mathcal{M}^{t}{}_{r}=0\,,\quad\mathcal{M}^{t}{}_{t}=-\rho\,,\quad \mathcal{M}^{r}{}_{r}=p_r\,,&\nonumber\\ &\mathcal{M}^{\theta}{}_{\theta}=p_t\,,\quad \mathcal{C}_r=0\,.&
\label{eq:reducedsystem}
\eea

We have not included $\mathcal{C}_t=0$, since the equation $\mathcal{M}^{t}{}_{r}=0$ implies either $Q'(r)f_{QQ}=0$
or $k-2cd\alpha=0$. In either case, the temporal component of the connection equation is automatically satisfied. The diagonal metric equations determine the matter content $(\rho,p_r,p_t)$ for a given $f(Q)$ and connection profile $\beta(r)$, while $\mathcal{C}_r=0$ provides the dynamical equation governing $\beta(r)$ itself.

The role of the quantity $\psi$ defined in Eq.~\eqref{psidef} is now transparent. In the general equations~\eqref{mtteq}--\eqref{mthetheeq} and~\eqref{b6}, the terms proportional to $\psi$ encode the coupling between $\alpha$ and the geometry. Once the regularity condition $1+c\,\alpha=0$ is imposed (see Sec.~\ref{sec:conn2sol}), $\psi$ vanishes identically, and the system simplifies considerably: the metric equations lose their explicit $\alpha$-dependence, and the connection profile is governed entirely by $\beta(r)$ and the constants $c$ and $k$. In this limit, the diagonal field equations depend on $\beta$ only through $Q(r)$ and $Q'(r)$, and the equation $\mathcal{C}_r=0$ becomes the sole nontrivial differential equation for the connection sector.

\section{Nontrivial solutions for connection I}\label{sec:conn1sol}

In this section, we investigate the solutions to the field equations associated with Connection~I, introduced in the previous section (see Eqs.~\eqref{eq:mtrconn1}--\eqref{eq:Cr_I}). To this end, we begin by solving Eq.~\eqref{eq:mtrconn1}. Since the constant $c$ is necessarily nonzero (cf.\ Sec.~\ref{sec:conn1}), this equation implies either $Q'(r)=0$ or $f_{QQ}=0$.

If $f_{QQ}=0$, then $f(Q)=a_0+a_1,Q$, where $a_0$ and $a_1$ are constants. This choice reduces the theory to GR with a cosmological constant $\Lambda=a_0/2$, and therefore does not lead to any new phenomenology.

Nevertheless, Eq.~\eqref{eq:mtrconn1} also admits solutions for a general $f(Q)$,  provided that the non-metricity scalar is constant, $Q(r)=Q_c$. Substituting this condition into Eq.~\eqref{eq:Q_conI}, which defines the non-metricity scalar for Connection~I, yields a differential equation for $\alpha(r)$, whose solution is
\begin{equation}
\alpha(r)=\frac{(2 a^2 +
   r^2) Q_c+12}{9r}+\frac{3 c_1-2 a\; \tan^{-1}\left(\frac{\sqrt{r^2-a^2}}{a}\right)}{3r\sqrt{r^2-a^2}}\,,
\end{equation}
where $c_1$ is an integration constant and $\tan^{-1}$ denotes the arctangent function. At large $r$, we have $\alpha(r)\approx Q_c r/9$. Near the throat ($r=a$), if we set $c_1=0$, the solution is approximated as $\alpha(r)\approx\left(2+a^2Q_c\right)/3a+\mathcal{O}(r-a)$. As a result, $\alpha(r)$ remains regular at the wormhole throat, i.e., $r=a$, as long as $c_1$ vanishes.

Therefore, Connection I admits nontrivial solutions only for constant $Q$, since $c$ is strictly non-zero. This is in contrast to Connection II, where nontrivial solutions with a non-constant $Q$ may exist, as we shall see in the following section.

For a constant non-metricity scalar, the remaining field equations simplify considerably. The connection equations, Eqs.~\eqref{eq:Ct_I}--\eqref{eq:Cr_I}, are automatically satisfied since they depend only on derivatives of $Q$. Moreover, the terms proportional to \(f_{QQ}Q'\) vanish identically in the diagonal metric equations, Eqs.~\eqref{eq:mtteq-1-gen}--\eqref{eq:mthetheeq-1-gen}.

\subsection{Energy Conditions}\label{subsec:connecec1}

The discussion of the energy conditions requires the choice of $f(Q)$. In fact, when $Q(r)$ is a constant, the last terms of Eqs.~\eqref{eq:mtteq-1-gen}, \eqref{eq:mrreq-1-gen}, \eqref{eq:mthetheeq-1-gen} do not contribute. Therefore, the energy density $\rho$ and pressures $p_r$ and $p_t$ are completely determined as long as the functional form of $f(Q)$ and the constant value of $Q_c$ are given. More explicitly, we have
\begin{align}
\rho&=\Lambda_Q-\frac{f_Qa^2}{r^4}\,,\label{eq:rho_QConst}\\
p_r&=-\Lambda_Q-\frac{f_Qa^2}{r^4}\,,\label{eq:pr_QConst}\\
p_t&=-\Lambda_Q+\frac{f_Qa^2}{r^4}\,,\label{eq:pt_QConst}
\end{align}
where
\begin{equation}
\Lambda_Q\equiv \frac{f_QQ}{2}-\frac{f}{2}\;,\qquad\textrm{with }Q=Q_c\,.
\end{equation}

One can see that at a large $r$, the anisotropic fluid becomes isotropic and behaves like an effective cosmological constant $\Lambda_Q$. In addition, the quantities that control the NEC are given by
\begin{equation}
\rho+p_r=-\frac{2f_Qa^2}{r^4}\,,\qquad \rho+p_t=0\,. \label{eq-NECsw}
\end{equation}
Consequently, whether the NEC is satisfied or violated is determined solely by the sign of $f_Q$, which depends on the choice of $f(Q)$ in the theory as well as the sign of $Q_c$. To illustrate this, we consider two examples: a quadratic model and an exponential model, as discussed below.

\subsubsection{Quadratic model}

We first focus on the model:
\bea
f(Q)=\frac{1}{\kappa}Q+\frac{\alpha_c}{\kappa^2}Q^2\,. \label{quadraticfq}
\eea
This choice is inspired by the results of Ref.~\cite{Subramaniam:2023okn}, which demonstrate that this function allows for NEC satisfaction in cosmological scenarios.  We therefore explore whether analogous results hold in the case of a wormhole geometry.

The effective cosmological constant, $\Lambda_Q$, for this model is 
\begin{equation}
\Lambda_Q=\frac{\alpha_c}{2\kappa^2}Q_c^2\,.
\end{equation}
In addition, according to Eq.~\eqref{eq-NECsw}, we have
\begin{equation}
\rho+p_r=-\frac{2a^2}{\kappa r^4}\left(1+\frac{2\alpha_c}{\kappa}Q_c\right)\,.\label{NECquadratic}
\end{equation}
Therefore, in the quadratic model, the NEC can be satisfied if $\alpha_c Q_c<-\kappa/2$, which can occur for a positive $\alpha_c$ if $Q_c$ is sufficiently negative. The fulfillment of WEC requires not only the fulfillment of NEC, but also a positive $\rho$ everywhere. This implies $Q_c^2 a^2\alpha_c>2(2 Q_c \alpha_c+\kappa)$ for a positive $\alpha_c$. 

On the other hand, a negative $\alpha_c$ requires a positive $Q_c$ to satisfy NEC. However,  a negative $\alpha_c$ implies a negative $\Lambda_Q$. Consequently, the WEC will not be satisfied in the whole spacetime, as can be checked from Eq.~\eqref{eq:rho_QConst}.

In Fig. \ref{fig:quadratic_conI}, we show an example with positive $\alpha_c$ for which both the NEC and WEC are satisfied throughout the entire spacetime. As can be seen, the three relevant quantities remain positive, with $\rho+p_t$ being identically zero. In addition, the energy density $\rho$ approaches a constant value in the asymptotic spatial region.

\begin{figure}
    \centering
    \includegraphics[width=1\linewidth]{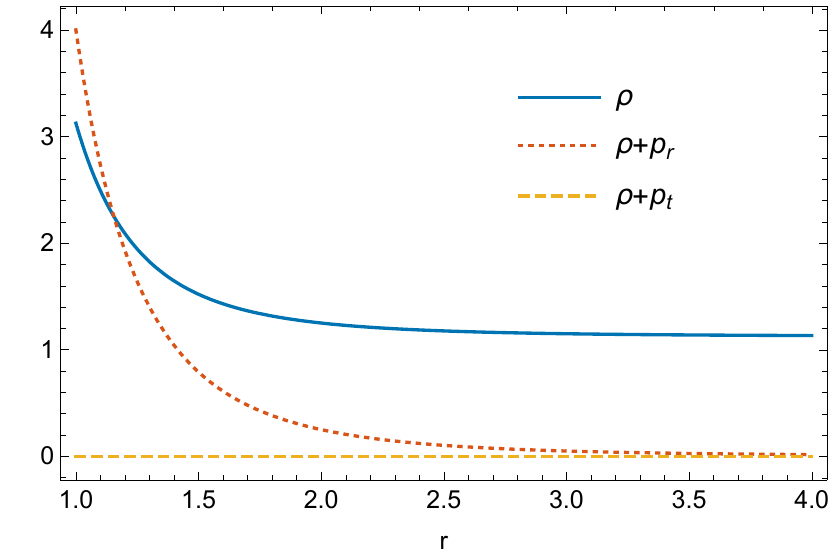}
    \caption{Profiles representing the energy conditions for Connection I, and the quadratic model, when $Q$ is constant and $\alpha_c$ is positive. The following values have been considered: $\kappa=\alpha_c=a=1$ and $Q_c=-1.5$. }
    \label{fig:quadratic_conI}
\end{figure}

We would like to emphasize that for the quadratic model, the effective cosmological constant $\Lambda_Q$ is proportional to $Q_c^2$. Therefore, the energy density and pressures are generically not zero at asymptotic infinity unless $Q_c$ is strictly zero. However, once $Q_c$ is chosen to be zero, the NEC is violated in the entire domain of $r$ (see Eq.~\eqref{NECquadratic}).

\subsubsection{Exponential model}

As the second example, we consider the exponential model
\begin{equation}
f(Q)=\gamma e^{-\frac{Q}{Q_0}}\,,\label{exponetialfQ}
\end{equation}
with $\gamma$ and $Q_0$ constant. In this case, we have
\begin{equation}
\Lambda_Q=-\frac{\gamma}{2}\left(1+\frac{Q_c}{Q_0}\right)e^{-\frac{Q_c}{Q_0}}\,,
\end{equation}
and
\begin{equation}
\rho+p_r=\frac{2a^2\gamma}{r^4Q_0}e^{-\frac{Q_c}{Q_0}}\,,\quad\rho=-p_t=\Lambda_Q+\frac{a^2\gamma}{r^4Q_0}e^{-\frac{Q_c}{Q_0}}\,.
\end{equation}
One can see that the NEC can be satisfied as long as $\gamma/Q_0>0$, irrespective of the sign of $Q_c$. Also, the effective cosmological constant $\Lambda_Q$ vanishes when we choose $Q_c=-Q_0$. Therefore, in the exponential model, one can find solutions that satisfy both NEC and WEC, and the anisotropic fluid asymptotically vanishes at large $r$. In this case, the fluid is always anisotropic, with energy density and pressures given by
\begin{equation}
\rho=p_r=-p_t=\frac{a^2\gamma}{r^4Q_0}e\; .
\end{equation}

\begin{figure}
    \centering
    \includegraphics[width=1\linewidth]{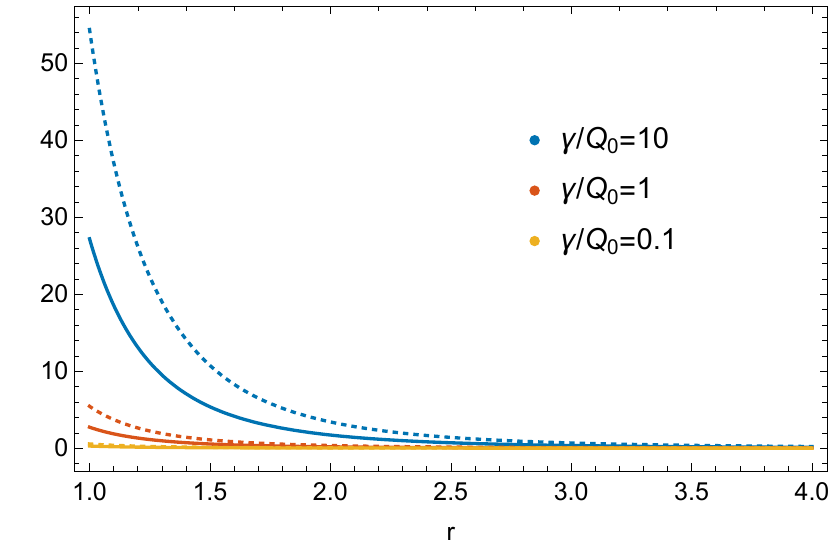}
    \caption{
    Profiles representing the energy conditions for Connection I, and the exponential model, when $Q$ is constant, fixing the throat at $a = 1$ and setting $\Lambda_Q = 0$ (which implies $Q_0 = -Q_c$), the profiles of $\rho$ and $\rho+p_r$ depend only on $\gamma / Q_0$. The solid lines represent the profile of $\rho(r)$, whereas the dashed lines are $\rho(r)+p_r(r)$ for the same coefficient of $\gamma / Q_0$.}
    \label{fig:rho+pr_exp}
\end{figure}

Fig.~\ref{fig:rho+pr_exp} shows the profiles of $\rho(r)$ and $\rho(r)+p_r(r)$ for the case $\Lambda_Q=0$ and several values of the ratio $\gamma/Q_0$. By contrast, the combination $\rho(r)+p_t(r)$ vanishes identically, independently of the value of this ratio.

\section{Nontrivial solutions for connection II}\label{sec:conn2sol}

In Sec.~\ref{sec:conn1sol}, we obtained the nontrivial wormhole solutions and the associated connection degrees of freedom for Connection I. We showed that for Connection~I, nontrivial solutions beyond GR only exist for a constant $Q$. In this section, we will consider Connection~II and discuss the associated nontrivial solutions. As we will show later, the richer mathematical structure of the field equations for Connection~II entails more possibilities of nontrivial solutions, allowing for solutions with not only a constant $Q$, but also a varying $Q$.

\subsection{Nonzero \texorpdfstring{$\beta(r)$ } <at the throat}\label{subsec:nonzerobeta}

In general, Connection~II and its associated dynamical degree of freedom are characterized by two free functions $\alpha(r)$ and $\beta(r)$. Suppose $Q(r)$ is not a constant, the field equation $\mathcal{M}^{t}{}_{r}=0$ and $\mathcal{C}_t=0$, i.e., Eqs.~\eqref{eq:mtrconn2} and \eqref{c_t}, require $\alpha(r)$ to be a constant satisfying $k-2cd\alpha=0$ for nontrivial solutions. Then, the connection degree of freedom is characterized by the function $\beta(r)$.

In this subsection, we will first show that the field equations of the theory already seem to constrain the behavior of $\beta(r)$ near the wormhole throat ($r\approx a$) for the equations to be satisfied. More explicitly, it turns out to be more convenient to assume $\beta(a)=0$ in order to satisfy the field equations.

Assuming $\beta(r)$ does not approach zero at the throat, i.e., $\beta(a)\ne0$, one can expand Eq.~\eqref{b7q} near the throat to get
\begin{equation}
Q(r)=\frac{\left(2+da^2c\right)\beta(a)}{2a^2\left(r-a\right)}+\mathcal{O}(r-a)^0\,,
\end{equation}
which shows that in general $Q(r)$ diverges at the throat. Then, to see whether $\mathcal{C}_r=0$ [Eq.~\eqref{b6}] can be satisfied or not, we need to choose a function $f(Q)$. We consider the quadratic model
\begin{equation}
f(Q)=\frac{1}{\kappa}Q+\frac{\alpha_c}{\kappa^2}Q^2\,,\label{quadraticmodel}
\end{equation}
then expand $\mathcal{C}_r$ near the throat. We get
\begin{equation}
\mathcal{C}_r=\frac{5\left(2+da^2c\right)^2\alpha_c\beta(a)^2}{8\kappa^2\sqrt{2a^3\left(r-a\right)^7}}+\mathcal{O}\left(r-a\right)^{-5/2}\,,
\end{equation}
which also diverges at $r=a$. Even for the special case where $2+da^2c=0$, we find that $\mathcal{C}_r$ diverges at the throat as $1/\sqrt{r-a}$ even though $Q(a)$ is finite. As a consequence, the field equation $\mathcal{C}_r=0$ cannot be satisfied at the throat in general if one has $\beta(a)\ne0$. Therefore, in the following discussions, we will focus mainly on the possibilities of looking for nontrivial solutions with $\beta(a)\rightarrow~0$ near the throat.

\subsection{\texorpdfstring{Constant $Q(r)=Q_c$}{beta(a)->0 and a constant Q(r)=Qc}}

In this subsection, we start with a simple setup with $Q(r)=Q_c$, i.e., $Q(r)$ is a constant. In this case, the field equations $\mathcal{M}^{t}{}_{r}=0$, $\mathcal{C}_t=0$, and $\mathcal{C}_r=0$ are trivially satisfied, regardless of the choice of $f(Q)$, and even without choosing $k-2cd\alpha=0$ in the beginning. In addition, according to the previous subsection, we will assume that $\beta(a)\rightarrow 0$. In order to avoid divergence in the connection components that results from a vanishing $\beta$ at the throat, we assume $1+c\alpha=0$, which implies that $\alpha(r)=-1/c$ is also a constant.

With a constant $Q(r)=Q_c$ and $\alpha(r)=-1/c$, an analytic and exact solution for $\beta(r)$ can be obtained from Eq.~\eqref{b7q}:
\begin{widetext}
\begin{equation}
\beta(r)=\frac{-2\left(r^2-a^2\right)\left(12+2a^2Q_c+Q_cr^2\right)+12a\sqrt{r^2-a^2}\tan^{-1}\left(\frac{\sqrt{r^2-a^2}}{a}\right)}{3r\left(4+d^2r^2+dkr^2\right)}+\frac{2\sqrt{r^2-a^2}C_1}{r\left(4+d^2r^2+dkr^2\right)}\,,\label{analyticexact1}
\end{equation}
\end{widetext}
for $4+d^2a^2+dka^2\ne0$, and 
\begin{widetext}
\begin{equation}
\beta(r)=\frac{a^2}{6r}\left[12+2a^2Q_c+Q_cr^2-\frac{6a}{\sqrt{r^2-a^2}}\tan^{-1}\left(\frac{\sqrt{r^2-a^2}}{a}\right)\right]+\frac{C_2}{r\sqrt{r^2-a^2}}\,,\label{analyticexact2}
\end{equation}
\end{widetext}
for $4+d^2a^2+dka^2=0$, where $C_1$ and $C_2$ are integration constants. It should be emphasized that these expressions do not depend on the choice of $f(Q)$ because Eq.~\eqref{b7q} does not depend on that. 

Expanding the solutions near the throat, we get
\begin{align}
\beta(r)=\frac{\sqrt{8(r-a)}C_1}{\sqrt{a}\left(4+d^2a^2+dka^2\right)}&-\frac{4\left(2+a^2Q_c\right)\left(r-a\right)}{4+d^2a^2+dka^2}\nonumber\\&+\mathcal{O}(r-a)^{3/2}\,,\label{asym1throat}
\end{align}
for $4+d^2a^2+dka^2\ne0$, and
\begin{equation}
\beta(r)=\frac{C_2}{\sqrt{2a^3\left(r-a\right)}}+\left(a+\frac{a^3Q_c}{2}\right)+\mathcal{O}(r-a)^{1/2}\,,\label{asym2throat}
\end{equation}
for $4+d^2a^2+dka^2=0$. For the first branch \eqref{asym1throat}, the requirement $\beta(a)=0$ is satisfied. We then further assume that the derivatives of $\beta(r)$ at the throat are also well defined, which requires $C_1=0$. As a result, we get
\begin{equation}
\beta(r)=
\begin{cases}
-\frac{4\left(2+a^2Q_c\right)\left(r-a\right)}{4+d^2a^2+dka^2}+\mathcal{O}(r-a)^{2}\qquad &r\rightarrow a\,,\\-\frac{2Q_cr}{3d(d+k)}+\mathcal{O}(r^{-1})\qquad &r\rightarrow\infty\,.\label{bran1exp}
\end{cases}
\end{equation}
In addition, according to the solution \eqref{analyticexact1}, $\beta(r)$ would diverge at $r=2/\sqrt{|d(d+k)|}$. Therefore, we have to assume $a>2/\sqrt{|d(d+k)|}$ in the parameter space such that the divergence of $\beta(r)$ does not appear outside the throat.

For the second branch \eqref{asym2throat} with $4+d^2a^2+dka^2=0$, the requirement $\beta(a)=0$ implies that $C_2=0$ and $Q_c=-2/a^2$. We then have
\begin{equation}
\beta(r)=
\begin{cases}
-\frac{4}{5a}(r-a)^2+\mathcal{O}(r-a)^{3}\qquad &r\rightarrow a\,,\\-\frac{r}{3}+\frac{4a^2}{3r}+\mathcal{O}(r^{-2})\qquad &r\rightarrow\infty\,.\label{bran2exp}
\end{cases}
\end{equation}
Note that once $Q_c=-2/a^2$, the solution \eqref{analyticexact1} with $C_1=0$ and $4+d^2a^2+dka^2=0$ matches exactly the solution \eqref{analyticexact2} with $C_2=0$.

In Fig.~\ref{fig:betaexact}, we choose some specific values of the parameters and show the exact solutions for $\beta(r)$. The black curves represent the first branch \eqref{analyticexact1} with $C_1=0$, $d=-k/2$, and $|k|=5/a$, with different values of $Q_c$. Note that in such a choice of parameters, we have $2/\sqrt{|d(d+k)|}=4a/5<a$, so no divergence occurs outside the throat. The red curve represents the second branch \eqref{analyticexact2} with $C_2=0$ and $Q_c=-2/a^2$. One sees that all the solutions vanish at the throat, and are everywhere regular outside the throat.

Finally, the regularity condition at the throat of $\xi(r)$, which is given by Eq.~\eqref{xi2}, is also relevant to the regularity of connection components, thus further constrains the solution space. For the first branch, the regularity condition requires $a^2d^3 - 4 k (3 + a^2 Q_c) - d (4 + a^2 k^2 + 4a^2 Q_c)=0$. We highlight this particular solution by the blue curve in Fig.~\ref{fig:betaexact}. For the second branch, the regularity condition cannot be satisfied because $\xi(r)$ diverges as $1/(r-a)^2$ near the throat.

\begin{figure}
    \centering
    \includegraphics[width=0.8\linewidth]{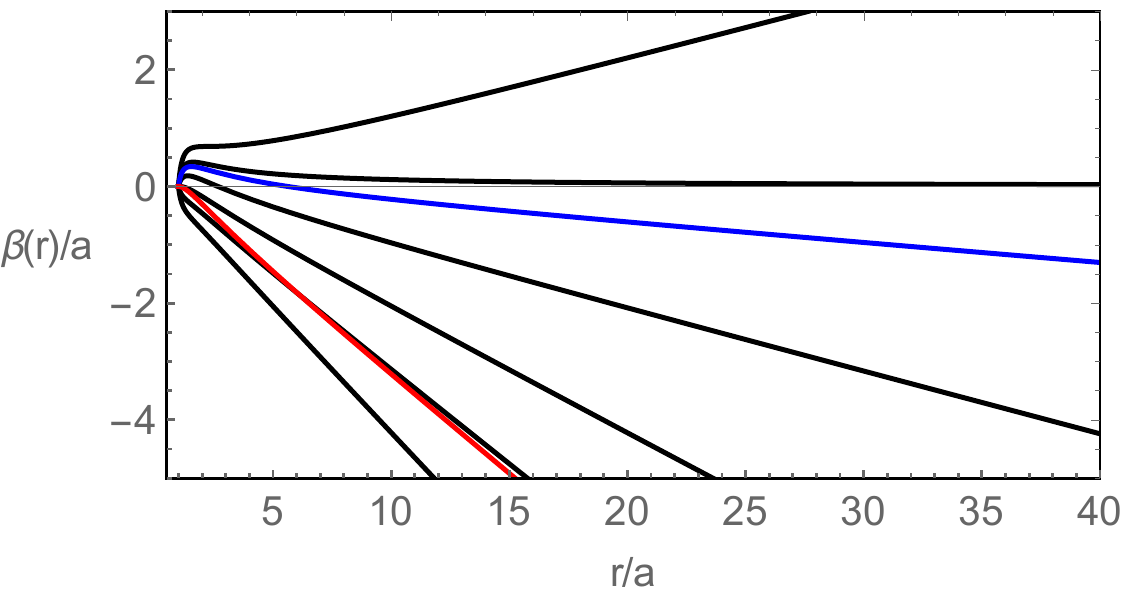}
    \caption{The solutions $\beta(r)$ for the first branch \eqref{analyticexact1} with $C_1=0$, $d=-k/2$, and $|k|=5/a$ (black, from top to bottom with $a^2Q_c=1$, $0$, $-1$, $-2$, $-3$, and $-4$), and for the second branch \eqref{analyticexact2} with $C_2=0$ and $Q_c=-2/a^2$ (red). A special curve in blue ($C_1=0$, $d=-k/2$, $|k|=5/a$, $a^2Q_c=-5/16$) represents the solution that satisfies the regularity condition of $\xi(r)$, i.e., Eq.~\eqref{xi2}, at the throat. All solutions extend to $r\rightarrow\infty$ without any divergence at finite $r$. Also, they all approach zero at the throat, and are all linear in $r$ at $r\rightarrow\infty$ as given by Eqs.~\eqref{bran1exp} and \eqref{bran2exp}, except for the one with $Q_c=0$, which decreases as $1/r$.}
    \label{fig:betaexact}
\end{figure}

\subsubsection{Energy conditions}

In the case of Connection II, the general expressions of the diagonal field equations, i.e., Eqs.~\eqref{mtteq}, \eqref{mrreq}, and \eqref{mthetheeq}, explicitly depend on $\beta(r)$. However, interestingly, under the assumption of a constant $Q$, the last terms of Eqs.~\eqref{mtteq}, \eqref{mrreq}, and \eqref{mthetheeq} do not contribute, and the diagonal field equations reduce exactly to those in the case of Connection I with constant $Q$, i.e., Eqs.~\eqref{eq:rho_QConst}, \eqref{eq:pr_QConst}, and \eqref{eq:pt_QConst}, where again   
\begin{equation}
\Lambda_Q\equiv \frac{f_QQ}{2}-\frac{f}{2}\qquad\textrm{with }Q=Q_c\,.
\end{equation}
Therefore, all the discussions and results obtained in Sec.~\ref{subsec:connecec1} are also applicable here. The only difference is that, for Connection I, the value of $Q_c$ can be arbitrarily chosen. However, for Connection II, the choice of $Q_c$ has to be subject to the constraint given by the regularity condition of $\xi(r)$ at the throat.

\subsubsection{Relaxing the assumption of a constant \texorpdfstring{$Q$}<}

At the beginning of this subsection, we directly assume a constant non-metricity scalar, i.e., $Q=Q_c$. This assumption significantly simplifies the field equations because $\mathcal{M}^{t}{}_{r}=0$, $\mathcal{C}_t=0$, and $\mathcal{C}_r=0$ are trivially satisfied. Before closing this subsection, we will show that such an assumption is actually not just for simplicity, but also stands for a natural choice in the sense that even if one relaxes the constant-$Q$ assumption, solving the field equations naturally leads to the branch of solutions with a constant $Q$, given a general assumption that $f_{QQ}$ does not diverge.

Based on the discussion in Sec.~\ref{subsec:nonzerobeta}, we assume that the function $\beta(r)$ approaches zero at the throat, i.e., $\beta(a)=0$. Then, we set $\alpha=-1/c$ such that $1+c\alpha=0$ in order to have well-defined connection components at the throat. Since the assumption of a constant $Q$ is relaxed, we need to impose the condition $k-2cd\alpha=0$ to ensure $\mathcal{M}^{t}{}_{r}=\mathcal{C}_t=0$. This implies $c=k/4$, or equivalently $d=-k/2$.

Assume the quadratic model{\footnote{The same analysis applies actually to all $f(Q)$ models in which $f_{QQ}$ does not diverge.}} \eqref{quadraticmodel} and consider the following series expansion of $\beta(r)$ near the throat $r=a$:
\begin{equation}
\beta(r)=\sum_{n=1}\beta_n(r-a)^n\,.
\end{equation}
Plugging it into Eqs.~\eqref{b6} and \eqref{b7q}, one can determine the expansion coefficients $\beta_n$ by expanding $\mathcal{C}_r$ near $r=a$ and taking $\mathcal{C}_r=0$ order by order in the expansion. Up to the second order in the expansion of $r-a$, we get
\begin{align}
\beta(r)&=\beta_1(r-a)-\frac{\left(16+11a^2k^2\right)\beta_1}{6a\left(a^2k^2-16\right)}\left(r-a\right)^2\nonumber\\
&+\mathcal{O}\left(r-a\right)^3\,,\label{f2}
\end{align}
with $ak\ne\pm4$, where $\beta_1$ remains undetermined. By taking Eq.~\eqref{f2} as the boundary condition for $\beta(r)$ near the throat and numerically solving Eq.~\eqref{b6}, we find that this solution is actually characterized by a constant $Q$ given by
\begin{equation}
Q(r)=\frac{\left(a^2k^2-16\right)\beta_1-32}{16a^2}\,. \label{Qrsol}
\end{equation}
In fact, this solution is essentially nothing but the analytic solution in Eq.~\eqref{analyticexact1} with $C_1=0$ and $d=-k/2$. The fact that $\beta_1$ remains undetermined corresponds to the fact that the constant $Q_c$ in Eq.~\eqref{analyticexact1} can be freely chosen before considering the regularity of $\xi(r)$. This can be confirmed by comparing the expansion coefficients in Eq.~\eqref{f2} and those in Eq.~\eqref{bran1exp}.

For the particular case in which  $ak=\pm 4$, one can repeat the same procedure and obtain
\bea
\beta(r)=-\frac{4}{5 a}(r-a)^2+\frac{8}{7a^2}(r-a)^3+\mathcal{O}\left(r-a\right)^4\,,\label{betaspecial}
\eea
near the throat. Using Eq.~\eqref{betaspecial} as the boundary condition at the throat and numerically integrating the field equations outward, we confirm that the solution has a constant $Q$:
\begin{equation}
Q(r)=-\frac{2}{a^2}\,.\label{specialQ}
\end{equation}
This solution is nothing but the analytic solution found in Eq.~\eqref{analyticexact2} with $C_2=0$ and $Q_c=-2/a^2$. The comparison between the expansion coefficients in Eq.~\eqref{betaspecial} and those in Eq.~\eqref{bran2exp} supports our argument.

Finally, we would like to emphasize that a constant $Q$ naturally satisfies $\mathcal{C}_r=0$ for a regular $f_{QQ}$. However, the analysis here does not exclude the possibility that other nontrivial solutions may appear when $f_{QQ}$ diverges at some finite $Q=Q_s$, and that the solution has a varying $Q(r)$. We will discuss this possibility in the next subsection.

\subsection{Solutions with varying \texorpdfstring{$Q$}<: Reconstruction method}\label{sec:reconstr}

In this subsection, we explore the possibility of constructing nontrivial solutions with a varying $Q$, but allowing $f_{QQ}$ to present divergences at some finite $Q$. Instead of considering a specific function $f(Q)$ to look for solutions, we will adopt a reconstruction method. More explicitly, we will specify the non-metricity $Q(r)$ in a way that some of its desired properties can be ensured in the first place. Then, we reconstruct the $f(Q)$ based on such a choice.

Following the discussions in the previous subsections, we again first impose $\alpha=-1/c$ and $c=k/4$. Then, after defining a new function $P(r)$:
\begin{equation}
P(r)\equiv Q'(r)f_{QQ}=\frac{df_Q(r)}{dr}\,,
\end{equation}
the connection field equation $\mathcal{C}_r=0$, in which $\mathcal{C}_r$ is given by Eq.~\eqref{b6}, can be written as an ordinary first-order differential equation of $\beta(r)$ (or $P(r)$) with a given $P(r)$ (or $\beta(r)$). This equation can be solved to get
\begin{equation}
P(r)\beta(r)^2=\frac{\sqrt{r^2-a^2}C_3}{r\left(16-k^2r^2\right)}\,,\label{Pgeneral}
\end{equation}
for $16-k^2a^2\ne0$, or
\begin{equation}
P(r)\beta(r)^2=\frac{C_4}{r\sqrt{r^2-a^2}}\,,\label{Pspecial}
\end{equation}
for $16-k^2a^2=0$, where $C_3$ and $C_4$ are integration constants. If the integration constants are zero, nontrivial solutions imply either $P(r)=0$, which are exactly the solutions we have discussed in the previous subsection, or $\beta(r)=0$, which are not consistent with the symmetry required for the connection components.{\footnote{According to Eq.~\eqref{xi2}, solutions with $\beta(r)=0$ have ill-defined connection components because of divergent $\xi(r)$.}   As a result, the integration constants have to be non-zero to look for nontrivial solutions with a varying $Q$.

To construct nontrivial solutions while keeping a physically reasonable configuration for the solutions, we expect that the non-metricity scalar $Q(r)$ approaches zero at $r\rightarrow\infty$. In addition, we expect that the function $f(Q)$ behaves as $Q$ in the same limit. In this regard, we consider a simple toy model with
\begin{equation}
Q(r)=\frac{p_1}{r}+\frac{p_2}{r^2}\,,\label{testQtrial}
\end{equation}
where $p_1$ and $p_2$ are constant. Inserting it into Eq.~\eqref{b7q}, one can obtain exact and analytic solutions for $\beta(r)$,
\begin{widetext}
\begin{equation}
\beta(r)=\frac{\sqrt{r^2-a^2}}{r\left(16-k^2r^2\right)}\left[C_5-4\left(8+p_1r+2p_2\right)\sqrt{r^2-a^2}+16a\tan^{-1}\left(\frac{\sqrt{r^2-a^2}}{a}\right)-4a^2p_1\cosh^{-1}\left(\frac{r}{a}\right)\right]\,,\label{betafinal}
\end{equation}
\end{widetext}
where $C_5$ is an integration constant. We note that to have a regular $\xi(r)$, we have to impose
\begin{equation}
C_5=0\,,\quad p_2 =\frac{ \left(3a^2k^2-80-16a p_1\right)}{16}\,.\label{xicondition}
\end{equation}
With these conditions, the expansions of $\beta(r)$ near the throat and at the asymptotic limit are given by
\begin{equation}
\beta(r)= 3(r-a)+\mathcal{O}(r-a)^2\,,\label{betavaryingqa}
\end{equation}
and
\begin{equation}
\beta(r)=\frac{4p_1}{k^2}+\mathcal{O}(1/r)\,,\label{betavaryingqlarge}
\end{equation}
respectively. Then, we can obtain $P(r)$ through Eq.~\eqref{Pgeneral} or \eqref{Pspecial}. Here, we show the large-$r$ expansion of $P(r)$:
\begin{equation}
P(r)=-\frac{C_3k^2}{16p_1^2r^2}+\mathcal{O}\left(r^{-3}\right)\,,
\end{equation}
for $16-k^2a^2\ne0$, and
\begin{equation}
P(r)=\frac{16C_4}{a^4p_1^2r^2}+\mathcal{O}\left(r^{-3}\right)\,,
\end{equation}
for $16-k^2a^2=0$.

Lastly, we integrate numerically $P(r)$ with respect to $r$ to get $f_Q(r)$. The boundary condition is chosen in the numerical integration at large $r$ such that $f_Q\rightarrow 1$ when $r\rightarrow\infty$. This can be achieved because $P(r)$ scales as $1/r^2$ at large $r$. As $Q(r)\approx p_1/r$, one generally obtains $f_Q\approx 1+c_0 Q$ at large $r$ with $c_0$ a coefficient determined by the parameters in the constructed solutions.

Before showing the explicit results, we would like to mention some caution that has to be taken into account in choosing the parameters. First, according to Eqs.~\eqref{betavaryingqa} and \eqref{betavaryingqlarge}, one can see that if $p_1<0$, the function $\beta(r)$ approaches a negative value at large $r$, meaning that $\beta(r)$ may either cross zero somewhere outside the throat, or diverge outside the throat when $k^2<16/a^2$. If $\beta(r)$ crosses zero outside the throat, the function $P(r)$, which is either given by Eq.~\eqref{Pgeneral} or Eq.~\eqref{Pspecial}, diverges there, and the solutions are not well-defined. On the other hand, if $\beta(r)$ diverges outside the throat, the regularity condition of connection components is violated. Therefore, in order to find valid solutions, we will take $p_1>0$. Second, we require the non-metricity scalar $Q(r)$ to be a monotonic function in $r\in(a,\infty)$. A monotonic $Q(r)$ in $r\in(a,\infty)$ is preferred in the reconstruction procedure because otherwise the solution $f_Q$ that we will reconstruct later as a function of $Q$ becomes a multivalued function. For the function $Q(r)$ chosen in Eq.~\eqref{testQtrial}, we then assume that the single turning point is located at the throat. This then fixes $p_1$ as
\begin{equation}
p_1=\frac{3a^2k^2-80}{8a}\,,\label{constrainp1}
\end{equation}
which means that $k^2>80/\left(3a^2\right)\approx26.67/a^2$ should be satisfied to have a positive $p_1$.

\begin{figure}
    \centering
    \includegraphics[width=1\linewidth]{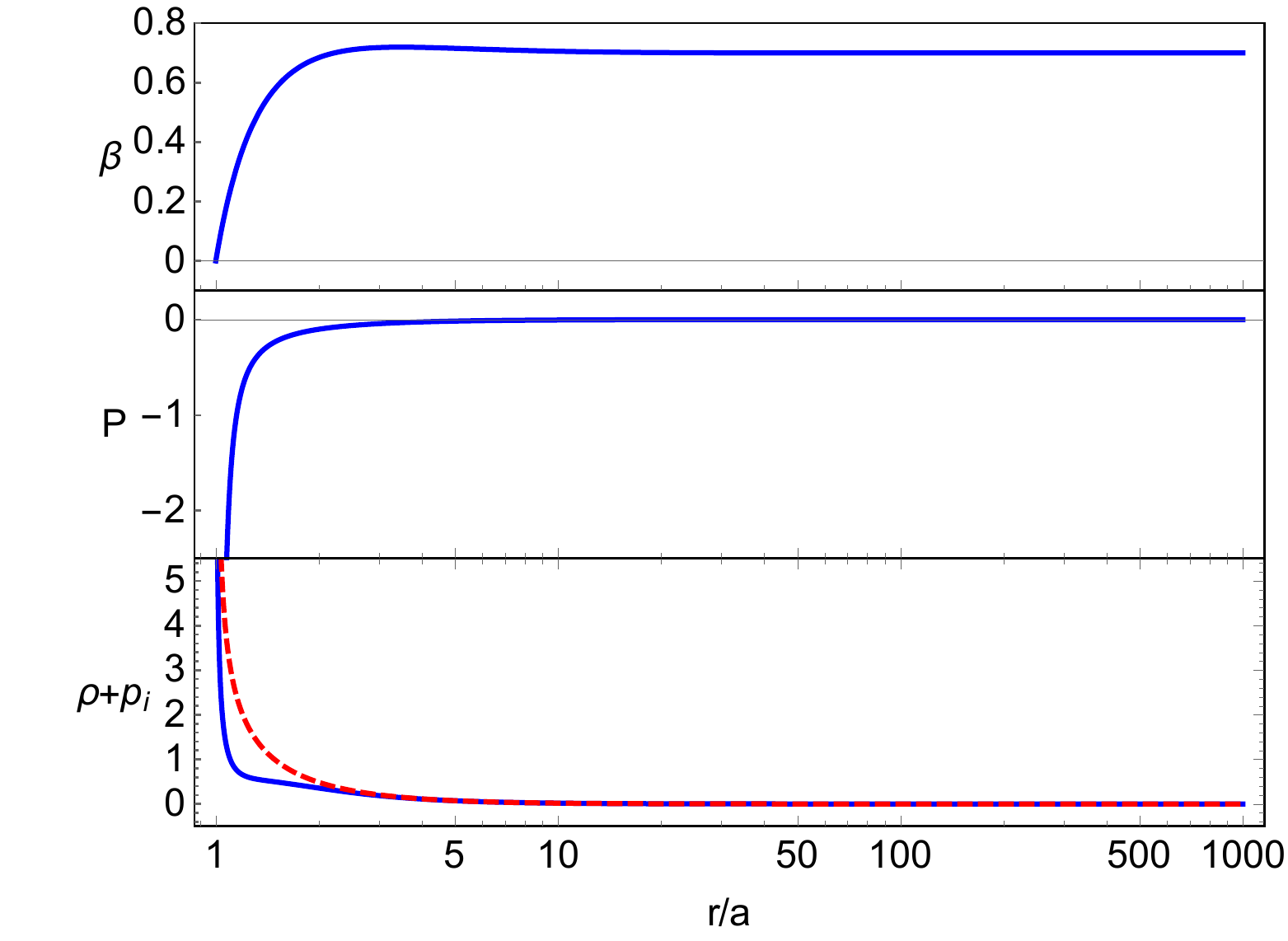}
    \caption{The nontrivial solutions reconstructed by setting $Q=p_1/r+p_2/r^2$. Here we take the solution branch of Eq.~\eqref{Pgeneral}. We choose $C_3=10$ and $k^2=50$ in the unit of $a$. For the value of $p_1$, we use Eq.~\eqref{constrainp1}. In the first two panels, we show $\beta(r)$ and $P(r)$. The NEC is satisfied for the entire region of $r\ge a$ as shown in the third panel, where $\rho+p_r$ and $\rho+p_t$ are depicted in blue and red, respectively.}
    \label{fig:varyingQsolution}
\end{figure}

For demonstration, in Fig.~\ref{fig:varyingQsolution}, we consider $Q(r)$ given in Eq.~\eqref{testQtrial}, then show a nontrivial solution obtained by taking the branch of Eq.~\eqref{Pgeneral} with $k^2=50/a^2$ and $p_1$ given by Eq.~\eqref{constrainp1}. The corresponding $\beta(r)$ given by Eq.~\eqref{betafinal} is shown in the first panel. Note that the conditions in Eq.~\eqref{xicondition} have to be satisfied as well. Then, using Eq.~\eqref{Pgeneral}, one can obtain $P(r)$, which is shown in the second panel. The NEC are satisfied for this particular solution, as shown in the last panel, where $\rho+p_r$ and $\rho+p_t$ are depicted in blue and red, respectively. We then reconstruct the function $f_Q$ as a function of $Q$ and show it in Fig.~\ref{fig:varyingQfq}. This can be done by parameterizing both $Q(r)$ and $f_Q(r)$ using the radial coordinate $r$. One can see that as $r\rightarrow\infty$, the non-metricity scalar $Q$ vanishes and $f_Q$ approaches unity at the leftmost point of the curve. Also, it turns out that $f_Q$ diverges at the throat, where $Q$ remains finite. In fact, we find that, near the throat, the function $f_Q$ scales as $f_Q\propto \left(Q_a-Q\right)^{-1/4}$, where $Q_a$ is the value of $Q$ at the throat. In summary, we show that nontrivial solutions that satisfy the NEC indeed exist even for varying $Q(r)$.

\begin{figure}
    \centering
    \includegraphics[width=1\linewidth]{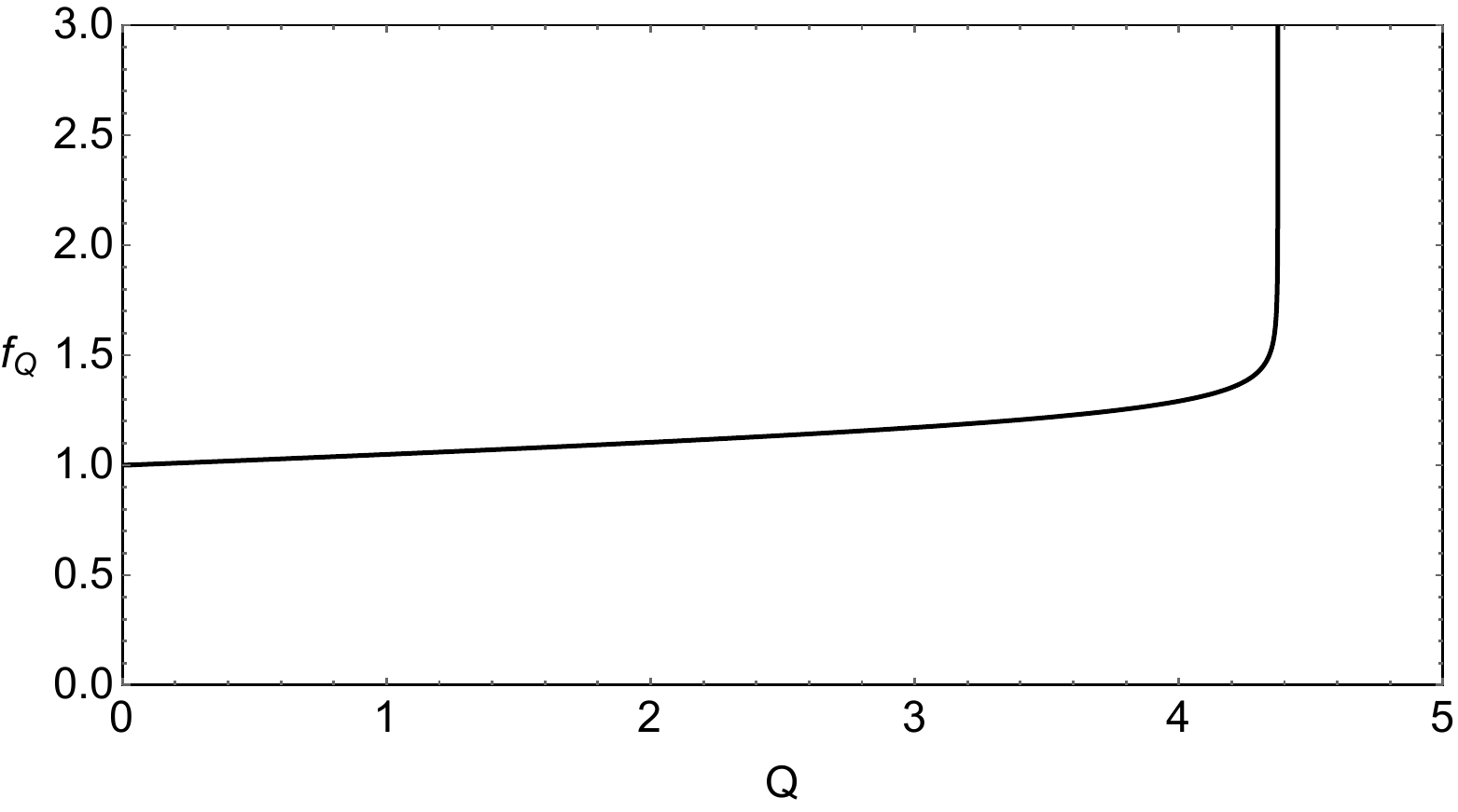}
    \caption{$f_Q$ as a function of $Q$ (in the unit of $a$) obtained by the solution shown in Fig.~\ref{fig:varyingQsolution}. The non-metricity scalar $Q$ vanishes at $r\rightarrow\infty$, where $f_Q$ approaches one.}
    \label{fig:varyingQfq}
\end{figure}

Finally, we would like to emphasize that, although the solutions constructed above are obtained by specifying $Q(r)$ first followed by solving $\beta(r)$, $P(r)$, and $f_Q$, one can also obtain solutions by specifying $P(r)$ first, with $\beta(r)$, $Q(r)$, and $f_Q$ reconstructed accordingly. However, from the practical point of view, in the latter approach, it turns out that the solution for $Q(r)$ obtained by solving Eq.~\eqref{b7q} may not be a monotonic function of $r$ within $r\in(a,\infty)$. As we have mentioned, a monotonic $Q(r)$ in $r\in(a,\infty)$ is preferred in the reconstruction procedure to ensure the reconstructed $f_Q$ as a function of $Q$ to be a single-valued function. In our approach demonstrated above, we can straightforwardly choose the function $Q(r)$ given by Eq.~\eqref{testQtrial} and the parameters $(p_1,p_2)$ in a way to ensure the monotonic behavior of $Q(r)$ in $r\in(a,\infty)$.

\Needspace{6\baselineskip} 
\section{Conclusions}\label{sec:conc}

In this paper, we have studied the Ellis--Bronnikov wormhole within the framework of $f(Q)$ modified gravity, with particular emphasis on the possibility of satisfying the energy conditions corresponding to the NEC and WEC. We have shown that the structure and symmetries of this spacetime, together with the requirements of symmetric teleparallel gravity, lead to two distinct possible affine connections.

We have then derived the field equations by varying the action with respect to the metric and the affine connection. These equations have been evaluated for the two distinct families of connections, assuming an anisotropic perfect fluid as the matter source.

For each connection, we thus obtain a system of equations consisting of one nontrivial off-diagonal metric equation, three diagonal metric equations determining the components of the energy-momentum tensor, and two nontrivial connection field equations.

For Connection~I, the solutions are restricted to either GR with a cosmological constant or a constant non-metricity scalar, $Q'(r)=0$, for a general $f(Q)$. Focusing on the latter case, the condition of constant $Q$ determines the form of the connection degree of freedom $\alpha(r)$. We identified specific forms of $f(Q)$ for which both the NEC and WEC are satisfied. In particular, we considered quadratic and exponential models, which mainly differ in their asymptotic behavior. The exponential model offers a qualitative advantage over the quadratic one: by choosing $Q_c = - Q_0$, the effective cosmological constant $\Lambda_Q$ vanishes, and the matter content decays to zero at spatial infinity. The quadratic model, by contrast, always produces a nonzero asymptotic energy density for $Q_c \neq 0$, which is the regime required by the NEC.

On the other hand, Connection~II exhibits a richer mathematical structure, allowing for nontrivial solutions with a non-constant $Q$. In this case, the connection is characterized by two free functions, $\alpha$ and $\beta$. Our first result for non-constant $Q$ is that setting $\beta(a)=0$ provides a convenient choice for satisfying the field equations. Consequently, we will adopt this assumption throughout the rest of the work, which implies that $\alpha$ is constant.

For Connection II, we have shown that, if $f_{QQ}$ is to remain finite, $Q$ must be constant. Consequently, $\alpha$ must also be constant when $\beta$ is required to vanish at the throat in order to avoid divergences. This allows us to determine the form of $\beta(r)$ for any choice of $f(Q)$ and identify the corresponding parameter space that avoids potential divergences. Interestingly, once $Q$ is taken to be constant, the equations associated with the matter sector become identical to those obtained for Connection~I. Therefore, the conclusions regarding the energy conditions are the same in both cases, with the only difference being that, for Connection~II, the value of $Q$ is constrained by the allowed parameter space.

However, for this connection, we can also allow for a non-constant $Q$ in order to depart as much as possible from the solutions obtained for Connection~I. This, however, comes at the cost of allowing $f_{QQ}$ to diverge at a finite value of $Q$. This  enables us to perform a reconstruction procedure for a given $Q(r)$ by appropriately choosing the parameter space. As a result, we have shown that nontrivial solutions satisfying the NEC also exist for a varying $Q(r)$.

It should be mentioned that the solutions we have found in this work all share the same wormhole geometry, i.e., the Ellis-Bronnikov wormhole, irrespective of the choice of connections or whether $Q$ is a constant or not. Therefore, from the observational probes that only rely on the computations at the level of the background metric, such as the motion of test particles, the solutions are not distinguishable from one another. They are not distinguishable from the Ellis-Bronnikov wormholes constructed in GR via such observational probes either. To break such degeneracy at the background level, one can consider gravitational perturbations of the wormhole. At the level of perturbations, the extra connection degrees of freedom may play some role in the dynamics in a way that differs among the solutions we have found, potentially allowing us to test them through, for instance, gravitational waves.

Another potential research direction is to adopt the reconstruction method we have used in Sec.~\ref{sec:reconstr} to study other compact objects in $f(Q)$ gravity. As we have shown in this work, the reconstruction method is suitable for exploring, in a systematic manner, possible solutions in non-GR branches that still preserve affine connection degrees of freedom. The solutions so obtained could be associated with a functional $f(Q)$ whose derivatives diverge at the configuration where non-GR effects become maximum. Such a property could help bypass the theoretical challenges outlined in Ref.~\cite{Heisenberg:2025xke} to find non-GR solutions, not just for neutron stars, but also for other scenarios, like regular black holes, black hole mimickers, boson stars, and other ultracompact objects. We will explore these directions elsewhere.

\section*{Acknowledgments}

IA is supported by the Basque Government Grant No. IT1628-22, by Grant No. PID2021-123226NB-I00 and No. PID2025-174372NB-I00 (funded by MCIN/AEI/10.13039/501100011033, and by the Grant JDC2023-052132-I (Juan de la Cierva) funded by MICIU/AEI/10.13039/501100011033 and by “ESF+”. M B-L is supported by the Basque Foundation for Science, Ikerbasque. Our work is supported by the Spanish Grant PID2023-149016NB-I00 funded by MCIN/\allowbreak AEI/\allowbreak 10.13039/\allowbreak 501100011033 and by the ERDF, “A way of making Europe”. It is also supported by Basque Government Grant No.~IT1628-22 (Spain). CYC is supported by the Special Postdoctoral Researcher (SPDR) Program at RIKEN and RIKEN Incentive Research Grant (Shoreikadai) 2025. K.D. is part of the BridgingCosmology project which is financed by Xjenza Malta and the Scientific and Technological Research Council of T\"{U}B\.{I}TAK, through the Xjenza Malta--T\"{U}B\.{I}TAK 2024 Joint Call for R\&I projects under Grant No.~124N627. This article is based upon work from COST Action CA21136 Addressing observational tensions in cosmology with systematics and fundamental physics (CosmoVerse) supported by COST (European Cooperation in Science and Technology).

\bibliography{Bib}

@article{Dehyadegari:2026vfd,
    author = "Dehyadegari, A. and Sheykhi, A.",
    title = "{New Black hole Solutions in $f(\mathbb{Q})$ Gravity}",
    eprint = "2602.18732",
    archivePrefix = "arXiv",
    primaryClass = "gr-qc",
    month = "2",
    year = "2026"
}

@article{Zhang:2025tsv,
    author = "Zhang, Zhen-Xiao and Lan, Chen and Miao, Yan-Gang",
    title = "{Comment on ''Black holes in $f(\mathbb{Q})$ gravity''}",
    eprint = "2508.12912",
    archivePrefix = "arXiv",
    primaryClass = "gr-qc",
    month = "8",
    year = "2025"
}

@article{Bahamonde:2022zgj,
    author = {Bahamonde, Sebastian and J{\"a}rv, Laur},
    title = "{Coincident gauge for static spherical field configurations in symmetric teleparallel gravity}",
    eprint = "2208.01872",
    archivePrefix = "arXiv",
    primaryClass = "gr-qc",
    doi = "10.1140/epjc/s10052-022-10922-9",
    journal = "Eur. Phys. J. C",
    volume = "82",
    number = "10",
    pages = "963",
    year = "2022"
}

@article{Lin:2021uqa,
    author = "Lin, Rui-Hui and Zhai, Xiang-Hua",
    title = "{Spherically symmetric configuration in $f(Q)$ gravity}",
    eprint = "2105.01484",
    archivePrefix = "arXiv",
    primaryClass = "gr-qc",
    doi = "10.1103/PhysRevD.103.124001",
    journal = "Phys. Rev. D",
    volume = "103",
    number = "12",
    pages = "124001",
    year = "2021",
    note = "[Erratum: Phys.Rev.D 106, 069902 (2022)]"
}

@article{Zhao:2021zab,
    author = "Zhao, Dehao",
    title = "{Covariant formulation of f(Q) theory}",
    eprint = "2104.02483",
    archivePrefix = "arXiv",
    primaryClass = "gr-qc",
    reportNumber = "USTC-ICTS/PCFT-21-16",
    doi = "10.1140/epjc/s10052-022-10266-4",
    journal = "Eur. Phys. J. C",
    volume = "82",
    number = "4",
    pages = "303",
    year = "2022"
}

@article{Heisenberg:2025xke,
    author = "Heisenberg, Lavinia and Pastor-Marcos, Carlos",
    title = "{Neutron stars in f(Q) gravity}",
    eprint = "2512.03037",
    archivePrefix = "arXiv",
    primaryClass = "gr-qc",
    doi = "10.1103/dy3t-1sq3",
    journal = "Phys. Rev. D",
    volume = "114",
    number = "2",
    pages = "024019",
    year = "2026"
}

@article{Bouhmadi-Lopez:2018sto,
    author = "Bouhmadi-L{\'o}pez, Mariam and Chen, Che-Yu and Chen, Pisin and Yeom, Dong-han",
    title = "{Regular Instantons in the Eddington-inspired-Born-Infeld Gravity: Lorentzian Wormholes from Bubble Nucleations}",
    eprint = "1809.06579",
    archivePrefix = "arXiv",
    primaryClass = "gr-qc",
    doi = "10.1088/1475-7516/2018/10/056",
    journal = "JCAP",
    volume = "10",
    pages = "056",
    year = "2018"
}

@article{DAmbrosio:2021zpm,
    author = "D'Ambrosio, Fabio and Fell, Shaun D. B. and Heisenberg, Lavinia and Kuhn, Simon",
    title = "{Black holes in f(Q) gravity}",
    eprint = "2109.03174",
    archivePrefix = "arXiv",
    primaryClass = "gr-qc",
    doi = "10.1103/PhysRevD.105.024042",
    journal = "Phys. Rev. D",
    volume = "105",
    number = "2",
    pages = "024042",
    year = "2022"
}

@article{Hochberg:1997wp,
    author = "Hochberg, David and Visser, Matt",
    title = "{Geometric structure of the generic static traversable wormhole throat}",
    eprint = "gr-qc/9704082",
    archivePrefix = "arXiv",
    doi = "10.1103/PhysRevD.56.4745",
    journal = "Phys. Rev. D",
    volume = "56",
    pages = "4745--4755",
    year = "1997"
}

@article{Carroll:2003st,
    author = "Carroll, Sean M. and Hoffman, Mark and Trodden, Mark",
    title = "{Can the dark energy equation-of-state parameter $w$ be less than -1?}",
    eprint = "astro-ph/0301273",
    archivePrefix = "arXiv",
    reportNumber = "EFI-2003-01, SU-GP-03-1-1",
    doi = "10.1103/PhysRevD.68.023509",
    journal = "Phys. Rev. D",
    volume = "68",
    pages = "023509",
    year = "2003"
}

@article{Lobo:2005us,
    author = "Lobo, Francisco S. N.",
    title = "{Phantom energy traversable wormholes}",
    eprint = "gr-qc/0502099",
    archivePrefix = "arXiv",
    doi = "10.1103/PhysRevD.71.084011",
    journal = "Phys. Rev. D",
    volume = "71",
    pages = "084011",
    year = "2005"
}

@article{Cline:2003gs,
    author = "Cline, James M. and Jeon, Sangyong and Moore, Guy D.",
    title = "{The Phantom menaced: Constraints on low-energy effective ghosts}",
    eprint = "hep-ph/0311312",
    archivePrefix = "arXiv",
    reportNumber = "MCGILL-03-25",
    doi = "10.1103/PhysRevD.70.043543",
    journal = "Phys. Rev. D",
    volume = "70",
    pages = "043543",
    year = "2004"
}

@article{BeltranJimenez:2017tkd,
    author = "Beltr{\'a}n Jim{\'e}nez, Jose and Heisenberg, Lavinia and Koivisto, Tomi",
    title = "{Coincident General Relativity}",
    eprint = "1710.03116",
    archivePrefix = "arXiv",
    primaryClass = "gr-qc",
    reportNumber = "NORDITA-2017-100, IFT-UAM/CSIC-17-093, ITS-ETH-2017-10",
    doi = "10.1103/PhysRevD.98.044048",
    journal = "Phys. Rev. D",
    volume = "98",
    number = "4",
    pages = "044048",
    year = "2018"
}

@article{Nester:1998mp,
    author = "Nester, James M. and Yo, Hwei-Jang",
    title = "{Symmetric teleparallel general relativity}",
    eprint = "gr-qc/9809049",
    archivePrefix = "arXiv",
    reportNumber = "NCU-CCS-980904",
    journal = "Chin. J. Phys.",
    volume = "37",
    pages = "113",
    year = "1999"
}

@article{Heisenberg:2023lru,
    author = "Heisenberg, Lavinia",
    title = "{Review on f(Q) gravity}",
    eprint = "2309.15958",
    archivePrefix = "arXiv",
    primaryClass = "gr-qc",
    doi = "10.1016/j.physrep.2024.02.001",
    journal = "Phys. Rept.",
    volume = "1066",
    pages = "1--78",
    year = "2024"
}

@article{DAmbrosio:2021pnd,
    author = "D'Ambrosio, Fabio and Heisenberg, Lavinia and Kuhn, Simon",
    title = "{Revisiting cosmologies in teleparallelism}",
    eprint = "2109.04209",
    archivePrefix = "arXiv",
    primaryClass = "gr-qc",
    doi = "10.1088/1361-6382/ac3f99",
    journal = "Class. Quant. Grav.",
    volume = "39",
    number = "2",
    pages = "025013",
    year = "2022"
}

@article{Dimakis:2023uib,
    author = "Dimakis, N. and Roumeliotis, M. and Paliathanasis, A. and Christodoulakis, T.",
    title = "{Anisotropic solutions in symmetric teleparallel $f\left( Q\right) $-theory: Kantowski\textendash{}Sachs and Bianchi III LRS cosmologies}",
    eprint = "2304.04419",
    archivePrefix = "arXiv",
    primaryClass = "gr-qc",
    doi = "10.1140/epjc/s10052-023-11964-3",
    journal = "Eur. Phys. J. C",
    volume = "83",
    number = "9",
    pages = "794",
    year = "2023"
}

@book{bookLobo,
author = {Lobo, Francisco},
year = {2017},
month = {07},
pages = {},
title = {Wormholes, Warp Drives and Energy Conditions},
isbn = {978-3-319-55181-4},
doi = {10.1007/978-3-319-55182-1},
publisher={Springer International Publishing}
}

@article{Hochberg:1998ii,
    author = "Hochberg, David and Visser, Matt",
    title = "{The Null energy condition in dynamic wormholes}",
    eprint = "gr-qc/9802048",
    archivePrefix = "arXiv",
    reportNumber = "LAEFF-98-02",
    doi = "10.1103/PhysRevLett.81.746",
    journal = "Phys. Rev. Lett.",
    volume = "81",
    pages = "746--749",
    year = "1998"
}

@article{Hochberg:1998ha,
    author = "Hochberg, David and Visser, Matt",
    title = "{Dynamic wormholes, anti-trapped surfaces, and energy conditions}",
    eprint = "gr-qc/9802046",
    archivePrefix = "arXiv",
    reportNumber = "LAEFF-98-01",
    doi = "10.1103/PhysRevD.58.044021",
    journal = "Phys. Rev. D",
    volume = "58",
    pages = "044021",
    year = "1998"
}

@article{Kar:1995ss,
    author = "Kar, Sayan and Sahdev, Deshdeep",
    title = "{Evolving Lorentzian wormholes}",
    eprint = "gr-qc/9506094",
    archivePrefix = "arXiv",
    reportNumber = "IP-BBSR-95-38",
    doi = "10.1103/PhysRevD.53.722",
    journal = "Phys. Rev. D",
    volume = "53",
    pages = "722--730",
    year = "1996"
}

@article{Arellano:2006ex,
    author = "Arellano, Aaron V. B. and Lobo, Francisco S. N.",
    title = "{Evolving wormhole geometries within nonlinear electrodynamics}",
    eprint = "gr-qc/0608003",
    archivePrefix = "arXiv",
    doi = "10.1088/0264-9381/23/20/004",
    journal = "Class. Quant. Grav.",
    volume = "23",
    pages = "5811--5824",
    year = "2006"
}

@article{Lobo:2008zu,
    author = "Lobo, Francisco S. N.",
    title = "{General class of wormhole geometries in conformal Weyl gravity}",
    eprint = "0801.4401",
    archivePrefix = "arXiv",
    primaryClass = "gr-qc",
    doi = "10.1088/0264-9381/25/17/175006",
    journal = "Class. Quant. Grav.",
    volume = "25",
    pages = "175006",
    year = "2008"
}

@article{Ellis:1973yv,
    author = "Ellis, H. G.",
    title = "{Ether flow through a drainhole - a particle model in general relativity}",
    doi = "10.1063/1.1666161",
    journal = "J. Math. Phys.",
    volume = "14",
    pages = "104--118",
    year = "1973"
}

@article{Bronnikov:1973fh,
    author = "Bronnikov, K. A.",
    title = "{Scalar-tensor theory and scalar charge}",
    journal = "Acta Phys. Polon. B",
    volume = "4",
    pages = "251--266",
    year = "1973"
}

@article{Morris:1988cz,
    author = "Morris, M. S. and Thorne, K. S.",
    title = "{Wormholes in space-time and their use for interstellar travel: A tool for teaching general relativity}",
    doi = "10.1119/1.15620",
    journal = "Am. J. Phys.",
    volume = "56",
    pages = "395--412",
    year = "1988"
}

@book{Visser:1995cc,
    author = "Visser, Matt",
    title = "{Lorentzian wormholes: From Einstein to Hawking}",
    isbn = "978-1-56396-653-8",
    year = "1995",
    publisher="American Inst. of Physics"
}

@article{Lobo:2009ip,
    author = "Lobo, Francisco S. N. and Oliveira, Miguel A.",
    title = "{Wormhole geometries in f(R) modified theories of gravity}",
    eprint = "0909.5539",
    archivePrefix = "arXiv",
    primaryClass = "gr-qc",
    doi = "10.1103/PhysRevD.80.104012",
    journal = "Phys. Rev. D",
    volume = "80",
    pages = "104012",
    year = "2009"
}

@article{Boehmer:2012uyw,
    author = "Boehmer, Christian G. and Harko, Tiberiu and Lobo, Francisco S. N.",
    title = "{Wormhole geometries in modified teleparralel gravity and the energy conditions}",
    eprint = "1110.5756",
    archivePrefix = "arXiv",
    primaryClass = "gr-qc",
    doi = "10.1103/PhysRevD.85.044033",
    journal = "Phys. Rev. D",
    volume = "85",
    pages = "044033",
    year = "2012"
}

@article{Garcia:2010xb,
    author = "Garcia, Nadiezhda Montelongo and Lobo, Francisco S. N.",
    title = "{Wormhole geometries supported by a nonminimal curvature-matter coupling}",
    eprint = "1007.3040",
    archivePrefix = "arXiv",
    primaryClass = "gr-qc",
    doi = "10.1103/PhysRevD.82.104018",
    journal = "Phys. Rev. D",
    volume = "82",
    pages = "104018",
    year = "2010"
}

@article{MontelongoGarcia:2010xd,
    author = "Montelongo Garcia, Nadiezhda and Lobo, Francisco S. N.",
    title = "{Nonminimal curvature-matter coupled wormholes with matter satisfying the null energy condition}",
    eprint = "1012.2443",
    archivePrefix = "arXiv",
    primaryClass = "gr-qc",
    doi = "10.1088/0264-9381/28/8/085018",
    journal = "Class. Quant. Grav.",
    volume = "28",
    pages = "085018",
    year = "2011"
}

@article{Mehdizadeh:2015jra,
    author = "Mehdizadeh, Mohammad Reza and Kord Zangeneh, Mahdi and Lobo, Francisco S. N.",
    title = "{Einstein-Gauss-Bonnet traversable wormholes satisfying the weak energy condition}",
    eprint = "1501.04773",
    archivePrefix = "arXiv",
    primaryClass = "gr-qc",
    doi = "10.1103/PhysRevD.91.084004",
    journal = "Phys. Rev. D",
    volume = "91",
    number = "8",
    pages = "084004",
    year = "2015"
}

@article{KordZangeneh:2015dks,
    author = "Kord Zangeneh, Mahdi and Lobo, Francisco S. N. and Dehghani, Mohammad Hossein",
    title = "{Traversable wormholes satisfying the weak energy condition in third-order Lovelock gravity}",
    eprint = "1510.07089",
    archivePrefix = "arXiv",
    primaryClass = "gr-qc",
    doi = "10.1103/PhysRevD.92.124049",
    journal = "Phys. Rev. D",
    volume = "92",
    number = "12",
    pages = "124049",
    year = "2015"
}

@article{Rosa:2018jwp,
    author = "Rosa, Jo\~ao Lu\'\i{}s and Lemos, Jos\'e P. S. and Lobo, Francisco S. N.",
    title = "{Wormholes in generalized hybrid metric-Palatini gravity obeying the matter null energy condition everywhere}",
    eprint = "1808.08975",
    archivePrefix = "arXiv",
    primaryClass = "gr-qc",
    doi = "10.1103/PhysRevD.98.064054",
    journal = "Phys. Rev. D",
    volume = "98",
    number = "6",
    pages = "064054",
    year = "2018"
}

@article{Mol:2014ooa,
    author = "Mol, Igor",
    title = "{The Non-Metricity Formulation of General Relativity}",
    eprint = "1406.0737",
    archivePrefix = "arXiv",
    primaryClass = "gr-qc",
    doi = "10.1007/s00006-016-0749-8",
    journal = "Adv. Appl. Clifford Algebras",
    volume = "27",
    number = "3",
    pages = "2607--2638",
    year = "2017"
}

@article{Hehl:1994ue,
    author = "Hehl, Friedrich W. and McCrea, J.Dermott and Mielke, Eckehard W. and Ne'eman, Yuval",
    archivePrefix = "arXiv",
    doi = "10.1016/0370-1573(94)00111-F",
    eprint = "gr-qc/9402012",
    journal = "Phys. Rept.",
    pages = "1--171",
    reportNumber = "TAUP-N192-94, TAUP-192-94",
    title = "{Metric affine gauge theory of gravity: Field equations, Noether identities, world spinors, and breaking of dilation invariance}",
    volume = "258",
    year = "1995"
}

@article{Jarv:2018bgs,
    author = "Järv, Laur and Rünkla, Mihkel and Saal, Margus and Vilson, Ott",
    archivePrefix = "arXiv",
    doi = "10.1103/PhysRevD.97.124025",
    eprint = "1802.00492",
    journal = "Phys. Rev. D",
    number = "12",
    pages = "124025",
    primaryClass = "gr-qc",
    title = "{Nonmetricity formulation of general relativity and its scalar-tensor extension}",
    volume = "97",
    year = "2018"
}

@article{BeltranJimenez:2019esp,
    author = "Beltr{\'a}n Jim{\'e}nez, Jose and Heisenberg, Lavinia and Koivisto, Tomi S.",
    title = "{The Geometrical Trinity of Gravity}",
    eprint = "1903.06830",
    archivePrefix = "arXiv",
    primaryClass = "hep-th",
    doi = "10.3390/universe5070173",
    journal = "Universe",
    volume = "5",
    number = "7",
    pages = "173",
    year = "2019"
}

@article{Ayuso:2020dcu,
    author = "Ayuso, Ismael and Lazkoz, Ruth and Salzano, Vincenzo",
    title = "{Observational constraints on cosmological solutions of $f(Q)$ theories}",
    eprint = "2012.00046",
    archivePrefix = "arXiv",
    primaryClass = "astro-ph.CO",
    doi = "10.1103/PhysRevD.103.063505",
    journal = "Phys. Rev. D",
    volume = "103",
    number = "6",
    pages = "063505",
    year = "2021"
}

@article{BeltranJimenez:2019tme,
    author = "Beltr\'an Jim\'enez, Jose and Heisenberg, Lavinia and Koivisto, Tomi Sebastian and Pekar, Simon",
    title = "{Cosmology in $f(Q)$ geometry}",
    eprint = "1906.10027",
    archivePrefix = "arXiv",
    primaryClass = "gr-qc",
    doi = "10.1103/PhysRevD.101.103507",
    journal = "Phys. Rev. D",
    volume = "101",
    number = "10",
    pages = "103507",
    year = "2020"
}

@article{Subramaniam:2023okn,
    author = "Subramaniam, Ganesh and De, Avik and Loo, Tee-How and Goh, Yong Kheng",
    title = "{How Different Connections in Flat FLRW Geometry Impact Energy Conditions in f(Q)$f(Q)$ Theory?}",
    eprint = "2304.02300",
    archivePrefix = "arXiv",
    primaryClass = "gr-qc",
    doi = "10.1002/prop.202300038",
    journal = "Fortsch. Phys.",
    volume = "71",
    number = "8-9",
    pages = "2300038",
    year = "2023"
}

@article{BeltranJimenez:2017vop,
    author = "Beltr\'an Jim\'enez, Jose and Koivisto, Tomi S.",
    title = "{Modified Gravity with Vector Distortion and Cosmological Applications}",
    doi = "10.3390/universe3020047",
    journal = "Universe",
    volume = "3",
    number = "2",
    pages = "47",
    year = "2017"
}

@article{Frusciante:2021sio,
    author = "Frusciante, Noemi",
    title = "{Signatures of $f(Q)$-gravity in cosmology}",
    eprint = "2101.09242",
    archivePrefix = "arXiv",
    primaryClass = "astro-ph.CO",
    doi = "10.1103/PhysRevD.103.044021",
    journal = "Phys. Rev. D",
    volume = "103",
    number = "4",
    pages = "044021",
    year = "2021"
}

@article{Dimakis:2024fan,
    author = "Dimakis, Nikolaos and Terzis, Petros A. and Paliathanasis, Andronikos and Christodoulakis, Theodosios",
    title = "{Static, spherically symmetric solutions in f(Q)-gravity and in nonmetricity scalar-tensor theory}",
    eprint = "2410.04513",
    archivePrefix = "arXiv",
    primaryClass = "gr-qc",
    doi = "10.1016/j.jheap.2024.12.011",
    journal = "JHEAp",
    volume = "45",
    pages = "273--289",
    year = "2025"
}

@article{Dimakis:2025jrl,
    author = "Dimakis, Nikolaos and Giacomini, Alex and Paliathanasis, Andronikos and Panotopoulos, Grigorios",
    title = "{Relativistic stars in f(Q)-gravity: Exact analytic solution for the power-law case f(Q)=Q+bQ{\ensuremath{\nu}}}",
    eprint = "2503.14302",
    archivePrefix = "arXiv",
    primaryClass = "gr-qc",
    doi = "10.1016/j.aop.2026.170387",
    journal = "Annals Phys.",
    volume = "488",
    pages = "170387",
    year = "2026"
}

@article{Basilakos:2025olm,
    author = "Basilakos, Spyros and Paliathanasis, Andronikos and Saridakis, Emmanuel N.",
    title = "{Equivalence of f(Q) cosmology with quintom-like scenario: The phantom field as effective realization of the non-trivial connection}",
    eprint = "2503.19864",
    archivePrefix = "arXiv",
    primaryClass = "gr-qc",
    doi = "10.1016/j.physletb.2025.139658",
    journal = "Phys. Lett. B",
    volume = "868",
    pages = "139658",
    year = "2025"
}

@article{Dialektopoulos:2025ihc,
    author = "Dialektopoulos, Konstantinos F. and Leon, Genly and Paliathanasis, Andronikos",
    title = "{Noether symmetries in f(Q)-cosmology}",
    eprint = "2502.01202",
    archivePrefix = "arXiv",
    primaryClass = "gr-qc",
    doi = "10.1142/S0217732325501214",
    journal = "Mod. Phys. Lett. A",
    volume = "40",
    number = "29",
    pages = "2550121",
    year = "2025"
}

@article{Paliathanasis:2023pqp,
    author = "Paliathanasis, A. and Dimakis, N. and Christodoulakis, T.",
    title = "{Minisuperspace description of f(Q)-cosmology}",
    eprint = "2308.15207",
    archivePrefix = "arXiv",
    primaryClass = "gr-qc",
    doi = "10.1016/j.dark.2023.101410",
    journal = "Phys. Dark Univ.",
    volume = "43",
    pages = "101410",
    year = "2024"
}

@article{Paliathanasis:2024hvl,
    author = "Paliathanasis, Andronikos",
    title = "{Semi-Classical Limit and Quantum Corrections in Non-Coincidence Power-Law f(Q)-Cosmology}",
    eprint = "2409.07755",
    archivePrefix = "arXiv",
    primaryClass = "gr-qc",
    doi = "10.3390/axioms13090619",
    journal = "Axioms",
    volume = "13",
    number = "9",
    pages = "619",
    year = "2024"
}

@article{Shabani:2023xfn,
    author = "Shabani, Hamid and De, Avik and Loo, Tee-How and Saridakis, Emmanuel N.",
    title = "{Cosmology of f(Q) gravity in non-flat Universe}",
    eprint = "2306.13324",
    archivePrefix = "arXiv",
    primaryClass = "gr-qc",
    doi = "10.1140/epjc/s10052-024-12582-3",
    journal = "Eur. Phys. J. C",
    volume = "84",
    number = "3",
    pages = "285",
    year = "2024"
}

@article{Capozziello:2024lsz,
    author = "Capozziello, Salvatore and Shokri, Mehdi",
    title = "{Comparing inflationary models in extended Metric-Affine theories of gravity}",
    eprint = "2408.17415",
    archivePrefix = "arXiv",
    primaryClass = "gr-qc",
    doi = "10.1016/j.dark.2024.101698",
    journal = "Phys. Dark Univ.",
    volume = "46",
    pages = "101698",
    year = "2024"
}

@article{Capozziello:2022tvv,
    author = "Capozziello, Salvatore and Shokri, Mehdi",
    title = "{Slow-roll inflation in f(Q) non-metric gravity}",
    eprint = "2209.06670",
    archivePrefix = "arXiv",
    primaryClass = "gr-qc",
    doi = "10.1016/j.dark.2022.101113",
    journal = "Phys. Dark Univ.",
    volume = "37",
    pages = "101113",
    year = "2022"
}

@article{BeltranJimenez:2018vdo,
    author = "Beltr\'an Jim\'enez, Jose and Heisenberg, Lavinia and Koivisto, Tomi S.",
    title = "{Teleparallel Palatini theories}",
    eprint = "1803.10185",
    archivePrefix = "arXiv",
    primaryClass = "gr-qc",
    reportNumber = "NORDITA-2018-023, IFT-UAM/CSIC-18-035, IFT-UAM-CSIC-18-035",
    doi = "10.1088/1475-7516/2018/08/039",
    journal = "JCAP",
    volume = "08",
    pages = "039",
    year = "2018"
}

@article{Dong:2025pyz,
    author = "Dong, Yu-Qi and Lai, Xiao-Bin and Fan, Yu-Zhi and Liu, Yu-Xiao",
    title = "{Polarization modes of gravitational waves in general symmetric teleparallel gravity}",
    eprint = "2505.13298",
    archivePrefix = "arXiv",
    primaryClass = "gr-qc",
    month = "5",
    year = "2025"
}

@article{Lazkoz:2019sjl,
    author = "Lazkoz, Ruth and Lobo, Francisco S. N. and Ortiz-Ba\~nos, Mar\'\i{}a and Salzano, Vincenzo",
    title = "{Observational constraints of $f(Q)$ gravity}",
    eprint = "1907.13219",
    archivePrefix = "arXiv",
    primaryClass = "gr-qc",
    doi = "10.1103/PhysRevD.100.104027",
    journal = "Phys. Rev. D",
    volume = "100",
    number = "10",
    pages = "104027",
    year = "2019"
}

@article{Boiza:2025xpn,
    author = "Boiza, Carlos G. and Petronikolou, Maria and Bouhmadi-L{\'o}pez, Mariam and Saridakis, Emmanuel N.",
    title = "{Addressing H $_{0}$ and S $_{8}$ tensions within f(Q) cosmology}",
    eprint = "2505.18264",
    archivePrefix = "arXiv",
    primaryClass = "astro-ph.CO",
    doi = "10.1088/1475-7516/2025/12/011",
    journal = "JCAP",
    volume = "12",
    pages = "011",
    year = "2025"
}

@article{Chakraborty:2025qlv,
    author = "Chakraborty, Saikat and Dutta, Jibitesh and Gregoris, Daniele and Karwan, Khamphee and Khyllep, Wompherdeiki",
    title = "{Reproducing \ensuremath{\Lambda}CDM-like solutions in f(Q) gravity: a~comprehensive study across all connection branches}",
    eprint = "2501.15159",
    archivePrefix = "arXiv",
    primaryClass = "gr-qc",
    doi = "10.1088/1475-7516/2025/05/098",
    journal = "JCAP",
    volume = "05",
    pages = "098",
    year = "2025"
}

@article{Sahlu:2024pxk,
    author = "Sahlu, Shambel and de la Cruz-Dombriz, \'Alvaro and Abebe, Amare",
    title = "{Structure growth in f (Q) cosmology}",
    eprint = "2405.07361",
    archivePrefix = "arXiv",
    primaryClass = "gr-qc",
    doi = "10.1093/mnras/staf439",
    journal = "Mon. Not. Roy. Astron. Soc.",
    volume = "539",
    number = "2",
    pages = "690--703",
    year = "2025"
}

@article{Ayuso:2021vtj,
    author = "Ayuso, Ismael and Lazkoz, Ruth and Mimoso, Jos\'e Pedro",
    title = "{DGP and DGP-like cosmologies from f(Q) actions}",
    eprint = "2111.05061",
    archivePrefix = "arXiv",
    primaryClass = "astro-ph.CO",
    doi = "10.1103/PhysRevD.105.083534",
    journal = "Phys. Rev. D",
    volume = "105",
    number = "8",
    pages = "083534",
    year = "2022"
}

@article{Paliathanasis:2025hjw,
    author = "Paliathanasis, Andronikos",
    title = "{Testing non-coincident f(Q)-gravity with DESI DR2 BAO and GRBs}",
    eprint = "2504.11132",
    archivePrefix = "arXiv",
    primaryClass = "gr-qc",
    doi = "10.1016/j.dark.2025.101993",
    journal = "Phys. Dark Univ.",
    volume = "49",
    pages = "101993",
    year = "2025"
}

@article{Capozziello:2023vne,
    author = "Capozziello, Salvatore and De Falco, Vittorio and Ferrara, Carmen",
    title = "{The role of the boundary term in f(Q,~B) symmetric teleparallel gravity}",
    eprint = "2307.13280",
    archivePrefix = "arXiv",
    primaryClass = "gr-qc",
    doi = "10.1140/epjc/s10052-023-12072-y",
    journal = "Eur. Phys. J. C",
    volume = "83",
    number = "10",
    pages = "915",
    year = "2023"
}

@article{De:2023xua,
    author = "De, Avik and Loo, Tee-How and Saridakis, Emmanuel N.",
    title = "{Non-metricity with boundary terms: {\ensuremath{\mathsf{f}}}({\ensuremath{\mathsf{Q}}},{\ensuremath{\mathsf{C}}}) gravity and cosmology}",
    eprint = "2308.00652",
    archivePrefix = "arXiv",
    primaryClass = "gr-qc",
    doi = "10.1088/1475-7516/2024/03/050",
    journal = "JCAP",
    volume = "03",
    pages = "050",
    year = "2024"
}

@article{Ayuso:2026hqz,
    author = "Ayuso, Ismael",
    title = "{Degenerate and connection-dependent cosmological sectors in f(Q,C) gravity}",
    eprint = "2607.05521",
    archivePrefix = "arXiv",
    primaryClass = "gr-qc",
    month = "7",
    year = "2026"
}

@article{Artola:2025fup,
    author = "Artola, Mikel and Ayuso, Ismael and Lazkoz, Ruth and Olmo, Gonzalo and Salzano, Vincenzo",
    title = "{Geometric acceleration in f(Q, C) theories}",
    eprint = "2510.27415",
    archivePrefix = "arXiv",
    primaryClass = "gr-qc",
    doi = "10.1016/j.dark.2025.102180",
    journal = "Phys. Dark Univ.",
    volume = "50",
    pages = "102180",
    year = "2025"
}

@article{Ayuso:2025vkc,
    author = "Ayuso, Ismael and Bouhmadi-L{\'o}pez, Mariam and Chen, Che-Yu and Chew, Xiao Yan and Dialektopoulos, Konstantinos and Ong, Yen Chin",
    title = "{Insights in f(Q) cosmology: the relevance of the connection}",
    eprint = "2506.03506",
    archivePrefix = "arXiv",
    primaryClass = "gr-qc",
    reportNumber = "RIKEN-iTHEMS-Report-25",
    doi = "10.1088/1475-7516/2025/11/068",
    journal = "JCAP",
    volume = "11",
    pages = "068",
    year = "2025"
}

@article{Yang:2024tkw,
    author = "Yang, Yuhang and Ren, Xin and Wang, Bo and Cai, Yi-Fu and Saridakis, Emmanuel N.",
    title = "{Data reconstruction of the dynamical connection function in f(Q) cosmology}",
    eprint = "2404.12140",
    archivePrefix = "arXiv",
    primaryClass = "astro-ph.CO",
    doi = "10.1093/mnras/stae1905",
    journal = "Mon. Not. Roy. Astron. Soc.",
    volume = "533",
    number = "2",
    pages = "2232--2241",
    year = "2024"
}

@article{Dimakis:2022rkd,
    author = "Dimakis, N. and Paliathanasis, A. and Roumeliotis, M. and Christodoulakis, T.",
    title = "{FLRW solutions in f(Q) theory: The effect of using different connections}",
    eprint = "2205.04680",
    archivePrefix = "arXiv",
    primaryClass = "gr-qc",
    doi = "10.1103/PhysRevD.106.043509",
    journal = "Phys. Rev. D",
    volume = "106",
    number = "4",
    pages = "043509",
    year = "2022"
}

@article{Dimakis:2022wkj,
    author = "Dimakis, N. and Roumeliotis, M. and Paliathanasis, A. and Apostolopoulos, Pantelis S. and Christodoulakis, T.",
    title = "{Self-similar cosmological solutions in symmetric teleparallel theory: Friedmann-Lema{\^\i}tre-Robertson-Walker spacetimes}",
    eprint = "2210.10295",
    archivePrefix = "arXiv",
    primaryClass = "gr-qc",
    doi = "10.1103/PhysRevD.106.123516",
    journal = "Phys. Rev. D",
    volume = "106",
    number = "12",
    pages = "123516",
    year = "2022"
}

@article{Chen:2026lof,
    author = "Chen, Wen-Xiang and Zeeshan Gul, M.",
    title = "{Horizon Residues, Matter-Trace Winding and Test-Scalar Modes in f(Q,T) Gravity}",
    eprint = "2610.39370",
    archivePrefix = "arXiv",
    primaryClass = "gr-qc",
    month = "10",
    year = "2026"
}

\end{document}